\documentclass[prb,twocolumn,amsmath,amssymb,aps,longbibliography,superscriptaddress]{revtex4-2}

\usepackage{feynmp}
\usepackage{amsmath}
\usepackage{graphicx}
\usepackage{multirow}
\usepackage{latexsym}
\usepackage{textcomp}
\usepackage{verbatim}
\usepackage{color}
\usepackage{bm}
\usepackage{subfigure}
\usepackage{everysel}
\usepackage{keyval}
\usepackage{ragged2e}
\usepackage{dsfont}
\usepackage{amssymb}
\usepackage{enumitem}
\usepackage{pstricks}
\usepackage{mathtools}
\usepackage{kotex}

\usepackage{braket}
\usepackage{slashed} 
\usepackage{hyperref}        
\hypersetup{
     colorlinks=true,
     citecolor = red,    
     linkcolor=magenta,
     }

\usepackage{graphicx}
\usepackage{xcolor}
\usepackage{amsmath}
\usepackage{bbold}
\usepackage{amsfonts}
\usepackage{bbm}
\usepackage{comment}
\usepackage{orcidlink}

\newcommand{\osum}{{%
    \setbox0\hbox{\circ}%
    \rlap{\hbox to \wd0{\hss\sum\hss}}\box0
}}

\begin{document}

\title{Diverging entanglement of critical magnons in easy-axis antiferromagnets}

\author{Jongjun M. Lee\,\orcidlink{0000-0002-9786-1901}}
\affiliation{Department of Physics, Pohang University of Science and Technology (POSTECH), Pohang 37673, Korea}

\author{Hyun-Woo Lee\,\orcidlink{0000-0002-1648-8093}}
\thanks{Electronic Address: hwl@postech.ac.kr}
\affiliation{Department of Physics, Pohang University of Science and Technology (POSTECH), Pohang 37673, Korea}

\author{Myung-Joong Hwang\,\orcidlink{0000-0002-0176-6740}}
\thanks{Electronic Address: myungjoong.hwang@duke.edu}
\affiliation{Division of Natural and Applied Sciences, Duke Kunshan University, Kunshan, Jiangsu 215300, China}
\affiliation{Zu Chongzhi Center for Mathematics and Computational Science, Duke Kunshan University, Kunshan, Jiangsu 215300, China}

\begin{abstract}
We study the instability of antiferromagnets with easy-axis anisotropy under a magnetic field, uncovering single or even multiple phase transitions at the boundary between non-collinear and collinear spin orderings. Near the phase boundary, the entanglement between the sublattice magnons diverges due to the interplay among antiferromagnetic exchange interaction, anisotropy, and magnetic field. Furthermore, our study reveals that this magnetic criticality extends to a superradiant phase transition within cavity magnonics systems. The magnon-photon interaction results in diverging cavity photon numbers and squeezing in the ground state at the transition points between spin orderings. This investigation not only elucidates the criticality of multi-component squeezed magnons in antiferromagnets, but also proposes cavity photon measurements as a viable method for detecting magnetic phase transitions.
\end{abstract}
 
\date{\today}
\maketitle

\section{Introduction}
Antiferromagnets and ferrimagnets are characterized by staggered spin ordering from the spin exchange interaction~\cite{kittel1996introduction}. They are useful materials for spintronics for their rapid and efficient spin dynamics~\cite{jungwirth2016antiferromagnetic,nakata2017spin,baltz2018antiferro,kim2022ferrimagnetic}. Their spin excitations, known as magnons, exhibit unique attributes, such as wide frequency ranges~\cite{jungwirth2018multiple} and strong magnon-magnon coupling~\cite{liensberger2019exchange}. Recent studies have engineered magnon-magnon couplings by tilting spin ordering in synthetic magnets~\cite{shiota2020tunable,sud2020tunable,wang2024ultrastrong}. Superradiance of magnon has also been observed~\cite{bamba2022magnonic,kim2024observation}. Additionally, theoretical and experimental efforts have focused on integrating antiferromagnet and ferrimagnet into cavity magnonics setups to control and measure the quantum states of magnon~\cite{huebl2013high, tabuchi2014hybridizing, wang2019nonreciprocity,lachance2019hybrid, mandal2020coplanar, lachance2020entanglement, shim2020enhanced}.

\begin{figure}[t]
    \centering
    \includegraphics[width=\linewidth]{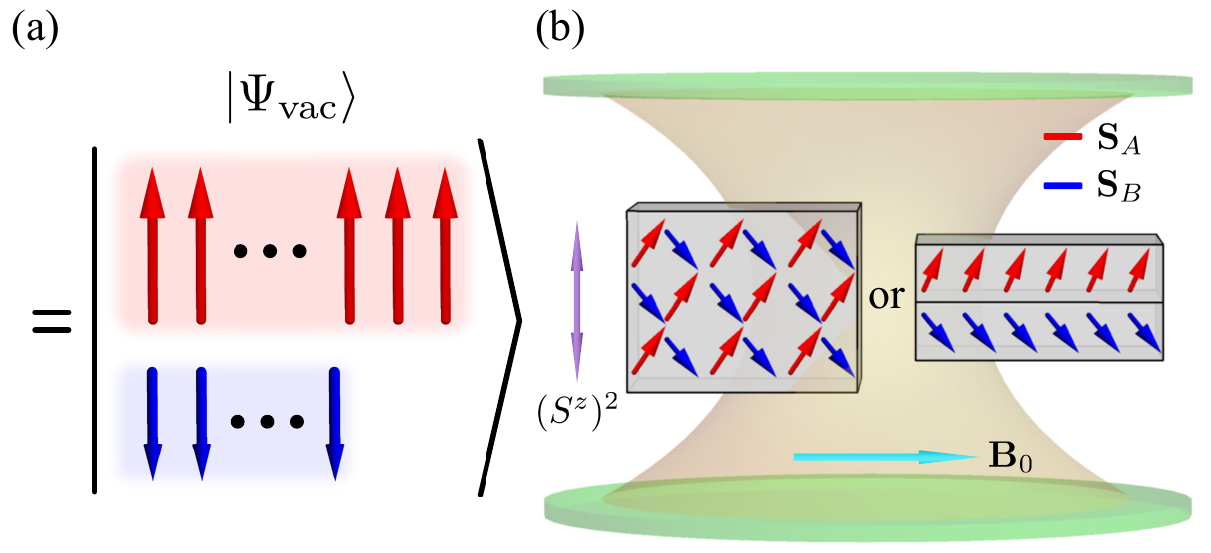}
    \caption{Schematic illustrations showing (a) the squeezed vacuum state of magnons and (b) antiferromagnets with spins $\mathbf{S}_{A, B}$. The antiferromagnets, influenced by an external magnetic field $\mathbf{B}_0$ perpendicular to the magnetic easy-axis anisotropy (proportional to $(S^{z})^{2}$), are placed inside the cavity. }
    \label{FIG1}
\end{figure}

An important characteristic of antiferromagnetic magnons is the two-mode squeezing due to the antiferromagnetic exchange interaction, which appears as magnon number-nonconserving terms in the Hamiltonian~\cite{zhao2004magnon,zhao2006magnon}. The two-mode squeezing interaction induces the entanglement between two different sublattice magnons [Fig.~\ref{FIG1}(a)]~\cite{kamra2017spin,kamra2019antiferromagnetic}, a genuine quantum mechanical nature of magnons that lacks a classical analogy~\cite{kamra2020magnon}. On the other hand, magnon squeezing is closely linked to a softening of magnon frequency and the onset of magnetic instability. A recent study on easy-axis ferromagnets has demonstrated that anisotropy can generate single-mode magnon squeezing of the Kittel mode, with the degree of squeezing controllable by an external magnetic field~\cite{zou2020tuning,lee2023cavity}. Interestingly, at a critical field strength, magnon squeezing diverges, leading to a magnetic phase transition. These findings suggest an intriguing possibility of controlling the entanglement between antiferromagnetic magnons by utilizing the anisotropy and an external magnetic field.  In particular, it would be interesting to explore whether a critical entanglement may occur and, if so, what types of spin orderings might result from magnetic instability. Devising an experimentally feasible scheme and enhancing coupling strengths with other degrees of freedom to measure such magnon entanglement are also important challenges to be addressed~\cite{lachance2019hybrid,lachance2020entanglement,lee2023topological,lee2023cavity,romling2023resolving,Romling2024}. These topics constitute the main focus of the current paper.

In this work, we theoretically study the physics of magnons in general antiferromagnetic materials in the presence of easy-axis anisotropy and an external magnetic field, which are oriented perpendicular to each other [Fig.~\ref{FIG1}(b)]. The interplay between the two-mode and single-mode squeezing effects due to the antiferromagnetic interaction and the anisotropy, respectively, drives the magnetic phase transition among the spin orderings, controlled by an external magnetic field. Notably, multiple magnetic phase transitions emerge in the ferrimagnets. We find that the magnon-magnon entanglement diverges at all critical points, demonstrating that the magnetic instability induces the critical antiferromagnetic magnon entanglement. Having established the theory of magnons in anisotropic antiferromagnets, we consider a cavity magnonics setup where the antiferromagnetic magnons are coupled to quantized cavity photons. We show that the spontaneous magnetization of the ferromagnets accompanies the cavity photon condensation, thereby translating the magnetic phase transitions due to the critical magnon entanglement to the superradiant phase transition~\cite{roman2021photon,liu2023switchable,lee2023cavity}. Moreover, both the quadrature squeezing and the number of photons diverge at the critical point. Measurement of cavity photons therefore provides information about magnonic entanglement and magnetic criticality in these systems.

This paper is organized as follows. In Sec.~\ref{Sec_Model}, we present a spin model for the antiferromagnet and the ferrimagnet. We present the analytic solution of the model in the case of zero magnetic field. In Sec.~\ref{Sec_OrderingsMagnons}, we solve the spin model and investigate the magnon properties in the ground state and the magnetic phase transitions in general conditions, considering the non-collinear ordering of the spin. In Sec.~\ref{Sec_CavityMagnonics}, we introduce the cavity and the magnon-photon coupling. We show the occurrence of the superradiant phase transition, investigate the physics of the magnon-polariton state, and discuss the experimental feasibility. In Sec.~\ref{Sec_Conclusion}, we provide a summary of our findings.

\section{Model} \label{Sec_Model}
We investigate a generic spin system featuring nearest-neighbor antiferromagnetic exchange interaction and magnetic easy-axis anisotropy, influenced by an external magnetic field on the cubic lattice. The anisotropy aligns with the $z$ axis, while the magnetic field lies along the $y$ direction, both mutually perpendicular. The corresponding spin Hamiltonian is expressed by,
\begin{equation}
\begin{aligned}
    H_0 = & J \sum_{\langle i,j\rangle } \mathbf{S}_{A,\mathbf{r}_{i} } \cdot \mathbf{S}_{B,\mathbf{r}_{j} } \\
         & - \sum_{i,\sigma} \left( K_{\sigma}  (S^{z}_{\sigma,\mathbf{r}_{i}})^{2}  + \gamma_{\sigma}B_{0} S^{y}_{\sigma,\mathbf{r}_{i}} \right),
\end{aligned}
\label{Eq_General_Ham}
\end{equation}
where $\langle i,j\rangle$ denotes the nearest neighbors. From now on, we define the convention $\sigma=A, B$ which denotes the sublattice and $\bar{\sigma}$ denotes another sublattice of $\sigma$. $J>0$ denotes the exchange interaction strength, $K_{\sigma}>0$ denotes the anisotropy strength, $\gamma_{\sigma}$ denotes the gyromagnetic ratio, and $B_{0}\ge 0$ is the magnetic field strength. We adopt $\hbar=1$ and symmetric gyromagnetic ratio $\gamma=\gamma_{\sigma}$ for simplicity. We note that this formulation applies to various types of magnets such as the synthetic antiferromagnet or synthetic ferrimagnet as depicted in Fig.~\ref{FIG1}(b)~\cite{grunberg1986layered,parkin1990oscillations,duine2018synthetic}.

\begin{figure*}[t!]
    \includegraphics[width=0.89\linewidth]{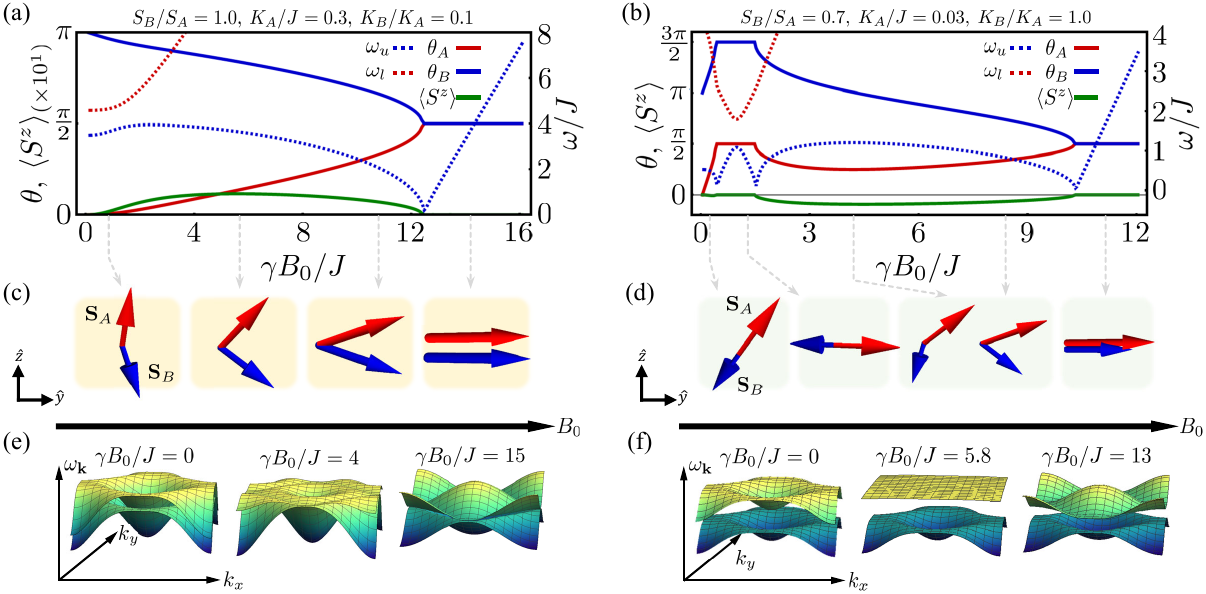}
    \caption{Spin ordering and magnon frequency spectra in different magnetic field strengths. Panels (a, c, e) depict characteristics in the antiferromagnetic regime, while panels (b, d, f) show corresponding features in the ferrimagnetic regime. (a, b) Plots of the mean-field solution $\theta_{\sigma}$, net spin angular momentum $\langle S^z \rangle$, and uniform magnon frequency $\omega_{u,l}$. (c, d) Schematics of the spin ordering. The gray dashed arrow below (a) and (b) indicates the corresponding magnetic field strength. (e, f) The frequency band of the magnon in momentum space.    }
    \label{Fig_Ordering_Magnon_1}
\end{figure*}

The model is amenable to analytical solutions in the absence of an external magnetic field ($B_{0}=0$)~\cite{kamra2019antiferromagnetic}. Under this condition, the mean-field solution is trivial: all spins align collinearly along the $z$ axis in an antiparallel configuration. We proceed by decomposing the spin Hamiltonian density into the magnon Hamiltonian density in momentum space along this trivial mean-field solution, employing the Holstein-Primakoff transformation as follows. 
\begin{equation}
\begin{aligned}
\mathcal{H}^{\rm t}_{\bf k} =& 
\begin{pmatrix}
    a^{\dagger}_{\mathbf{k}} & b_{-\mathbf{k}} 
\end{pmatrix}
\begin{pmatrix}
    f_{A} & g_{\mathbf{k}} \\ g_{\mathbf{k}} & f_{B}
\end{pmatrix}
\begin{pmatrix}
    a_{\mathbf{k}} \\ b^{\dagger}_{-\mathbf{k}} 
\end{pmatrix},
\end{aligned}
\end{equation}
where $a_{\mathbf{k}}$ and $b_{\mathbf{k}}$ are boson annihilation operators for magnons at the sites $A$ and $B$, $f_{A}=Jn_{1}S_{B}+2K_{A}S_{A}$, $f_{B}=Jn_{1}S_{A}+2K_{B}S_{B}$, $g_{\mathbf{k}}=-Jn_{1}\sqrt{S_{A}S_{B}}\gamma_{\mathbf{k}}$, $\gamma_{\mathbf{k}}=(1/n_{1})\sum_{\bm{\delta}}e^{i\mathbf{k}\cdot\bm{\delta}}$, $\bm{\delta}$ is a vector connecting the nearest neighbors, and $n_{1}$ is the coordination number. There are two-mode squeezing terms (off-diagonal terms) that create or annihilate two magnons across the sites simultaneously. These terms originate entirely from the antiferromagnetic exchange interaction. There are no squeezing terms within the same sublattices that reduce the uncertainty along $x$ or $y$, such as $a^{\dagger}_{\bf k}a^{\dagger}_{-\bf k}$ or $b^{\dagger}_{\bf k}b^{\dagger}_{-\bf k}$. This is because the mean-field solution possesses rotational symmetry about the $z$ axis, except when an external magnetic field is present. To diagonalize the Hamiltonian, we utilize the Bogoliubov transformation,
\begin{equation}
    \begin{pmatrix}
        a_{\bf k} \\ b^{\dagger}_{-\bf k}
    \end{pmatrix}
    =
    \begin{pmatrix}
        \cosh{r_{\bf k}} & -\sinh{r_{\bf k}} \\ -\sinh{r_{\bf k}} & \cosh{r_{\bf k}}
    \end{pmatrix}
    \begin{pmatrix}
        \alpha_{\bf k} \\ \beta^{\dagger}_{-\bf k}
    \end{pmatrix},
\end{equation}
where we refer to the transformed magnon as a squeezed magnon and $r_{\mathbf{k}}$, which is even in $\bf k$, represents the squeezing parameter. Subsequently, the Hamiltonian density in the diagonalized form is given by,
\begin{equation}
\begin{aligned}
\mathcal{H}^{\rm t}_{\bf k} =  \omega^{+}_{\mathbf{k}}\alpha^{\dagger}_{\mathbf{k}}\alpha_{\mathbf{k}} + \omega^{-}_{\mathbf{k}}\beta^{\dagger}_{\mathbf{k}}\beta_{\mathbf{k}},
\end{aligned}
\end{equation}
where the frequency of the eigenmode is
\begin{equation}
\omega^{\pm}_{\mathbf{k}} = \frac{ \pm(f_{A}-f_{B})+\sqrt{(f_{A}+f_{B})^{2}-4g^{2}_{\mathbf{k}}} }{2},    
\end{equation}
and the squeezing parameter is
\begin{equation}
    r_{\mathbf{k}} = \frac{1}{4} \log \Big( \frac{f_{A}+f_{B}-2g_{\mathbf{k}}}{f_{A}+f_{B}+2g_{\mathbf{k}}}\Big).
\end{equation} 
It is noteworthy that the frequency becomes degenerate if $f_{A}=f_{B}$ where the $\mathcal{P}\mathcal{T}$ symmetry of the mean-field spin solution is preserved. The squeezed magnons conserve their number and possess a spin angular momentum of $\pm 1$. The ground state is the squeezed vacuum state $|0\rangle_{\rm sq} = \prod_{\bf k} |0_{\bf k}\rangle_{\rm sq}$ where
\begin{equation}
|0_{\bf k}\rangle_{\rm sq}=\sum^{\infty}_{n_{\bf k}=0} \frac{(-\tanh r_{\bf k})^{n_{\bf k}}}{\cosh r_{\bf k}} | n_{\bf k},n_{\bf k} \rangle ,
\end{equation}
$\alpha_{\bf k}|0_{\bf k}\rangle_{\rm sq}=\beta_{\bf k}|0_{\bf k}\rangle_{\rm sq}=0$, and $|m_{\bf k},n_{\bf k}\rangle $ is the magnon Fock state with $n$ ($m$) nonsqueezed original magnons with momentum $\bf k$ at sublattice $A$ ($B$) [Fig.~\ref{FIG1}(a)]. Although the squeezed vacuum state contains no squeezed magnons, it consists of infinite Fock states of the original magnons. The squeezing term entangles the two original magnons with a nonzero entanglement entropy in the ground state; see Refs.~\cite{kamra2019antiferromagnetic,kamra2020magnon,wuhrer2022theory} for further details. Having reviewed the analytical solution in the absence of the magnetic field, we analyze in the following sections the intriguing roles of the Zeeman interaction due to both a classical magnetic field and quantized photons on the spin orderings and entanglements.

\section{Non-collinear Spin Orderings and Magnons} \label{Sec_OrderingsMagnons}
\subsection{Mean-field solutions}
We now investigate the spin Hamiltonian [Eq.~(\ref{Eq_General_Ham})] with more general parameter conditions in the presence of a magnetic field. As the magnetic field is turned on, the spin orientation no longer aligns with the easy axis but instead tilts towards directions between the easy axis and the field direction. We explore two distinct regimes: the antiferromagnetic regime, where the spin angular momenta $S_{\sigma}$ are symmetric ($S_{A} = S_{B}\equiv S$), and the ferrimagnetic regime, where they are asymmetric ($S_A\neq S_B$). To determine the spin orderings and their elementary excitations, our approach begins with calculating the mean-field theory of the spin Hamiltonian~\cite{rezende2019introduction}. We substitute the mean-field spin vector $\mathbf{S}_{\sigma}=S_{\sigma} (\sin\theta_{\sigma}\cos\phi_{\sigma},\:\sin\theta_{\sigma}\sin\phi_{\sigma},\:\cos\theta_{\sigma})$ into Eq.~(\ref{Eq_General_Ham}) and assume $\phi_{\sigma}=\pi/2$ for both sublattices. This assumption arises from a symmetry consisting of the product of time-reversal and mirror reflection along the $x$ axis. Then, the mean-field energy per spin is given by,
\begin{equation}
\begin{aligned}
\mathcal{E}_{\rm m} = &Jn_{1}S_{A}S_{B}\cos(\theta_{A}-\theta_{B})\\
&-\sum_{\sigma}(K_{\sigma}S^{2}_{\sigma}\cos^{2}{\theta_{\sigma}}+\gamma B_{0}S_{\sigma} \sin\theta_{\sigma} ).
\end{aligned}
\label{Eq_MF_1}
\end{equation}
We minimize this mean-field energy and calculate the minimization condition of $\{ \theta_{A},\: \theta_{B}\}$ which gives the spin orientation of the classical ground state. In the antiferromagnetic regime, if the anisotropy strengths are symmetric for both sublattices $(K=K_{\sigma})$, one can calculate analytical solutions with zero net spin angular momentum: $\theta_{\sigma}=\pi/2$ or
\begin{equation}
\sin\theta_{\sigma}=\frac{\gamma B_{0}}{2JSn_{1}+2KS}, \: (\theta_{A}\neq \theta_{B}).
\end{equation}
In the latter solution, $\theta_{\sigma}$ takes two distinct values, with $\theta_{A}$ for one sublattice and $\theta_{B}$ for the other. The former (latter) solution is stable if the right-hand side of the above equation is larger (lesser) than one. Therefore, there is a critical field strength $B_0^c\equiv(2JSn_1+2KS)/\gamma$ that separates two distinct phases. Note that the spin Hamiltonian possesses the mirror symmetry along the $y$ axis, $\mathcal{M}_y$, whose symmetry operation is given by
\begin{equation}
    S_{x} \xrightarrow[]{\mathcal{M}_{y}} -S_{x},\: S_{y}\xrightarrow[]{\mathcal{M}_{y}} +S_{y},\: S_{z}\xrightarrow[]{\mathcal{M}_{y}} -S_{z}.
\end{equation}
This operation is on-site and therefore does not interchange the sublattices. For $B_0>B_0^c$, the ground state respects the mirror symmetry. However, for $B_0<B_0^c$, this symmetry is spontaneously broken, resulting in two degenerate solutions corresponding to the exchange of two angles $\theta_\sigma$.

In the presence of asymmetry between the sublattices, the ground states are determined through numerical minimization. First, the mean-field solution for the antiferromagnetic regime is plotted in Fig.~\ref{Fig_Ordering_Magnon_1}(a), with schematic illustrations of the solution shown in (c). It demonstrates that the mirror symmetry-breaking transition persists for $K_A\neq K_B$. In the broken symmetry phase, the spin with smaller anisotropy strength rotates less than the spin with larger anisotropy strength at a given field strength, resulting in a finite net angular momentum and net magnetization, despite symmetric spin amplitudes; see $\langle S^z \rangle$ in Fig.~\ref{Fig_Ordering_Magnon_1}(a). These quantities vanish if the anisotropy strengths are identical ($K_{A}=K_{B}$). In the symmetric phase, a unique solution $\theta_\sigma=\pi/2$ exists with no net magnetization.

Second, the mean-field solution for the ferrimagnetic regime is plotted in Fig.~\ref{Fig_Ordering_Magnon_1}(b), with schematic illustrations shown in (d). Notably, four distinct phases with unique spin ordering emerge through consecutive phase transitions that break and restore mirror symmetry. Upon applying the magnetic field, the larger spin tilts toward the field direction while the smaller spin tilts away, resulting in antiparallel alignment between them. This collinear antiferromagnetic ordering, which breaks mirror symmetry, results in net spin angular momentum along the $z$ axis and double ground-state degeneracy. As the field strength increases, the larger (smaller) spin aligns parallel (antiparallel) to the field, specifically $\{\theta_{A},\theta_{B}\} = \{3\pi/2, \pi/2\}$. This configuration persists over a range of field strengths, $0.4 \lesssim B_{0} \lesssim 1.4$ in Fig.~\ref{Fig_Ordering_Magnon_1}(b), restoring mirror symmetry and resulting in no net spin angular momentum along the z-axis. Further increasing the field strength makes the Zeeman energy associated with the smaller spin pointing opposite to the field direction too large, causing it to start rotating toward the field direction. Meanwhile, the larger spin initially tilts away from the field to partially maintain antiferromagnetic ordering with the smaller spin but eventually rotates back to the field direction as the Zeeman interaction energy becomes the dominant energy scale. This results in a crossover from a non-collinear antiferromagnetic to ferromagnetic spin ordering within a broken-symmetry phase with non-zero net spin angular momentum. Finally, there is yet another phase transition to parallel spin ordering along the field direction, restoring mirror symmetry, as seen in the antiferromagnetic case. Our numerical analysis reveals a rich variety of spin orderings, including non-collinear antiferromagnetic order, in ferrimagnets. These distinctive properties, illustrated in Fig.~\ref{Fig_Ordering_Magnon_1}(b), emerge when the ratio of the two angular momentum amplitudes deviates significantly from one. For $S_{B}/S_{A}\simeq 1$, the mean-field solution closely resembles the antiferromagnetic regime, although this is not depicted in the figure.

\begin{figure*}[t!]
    \includegraphics[width=\linewidth]{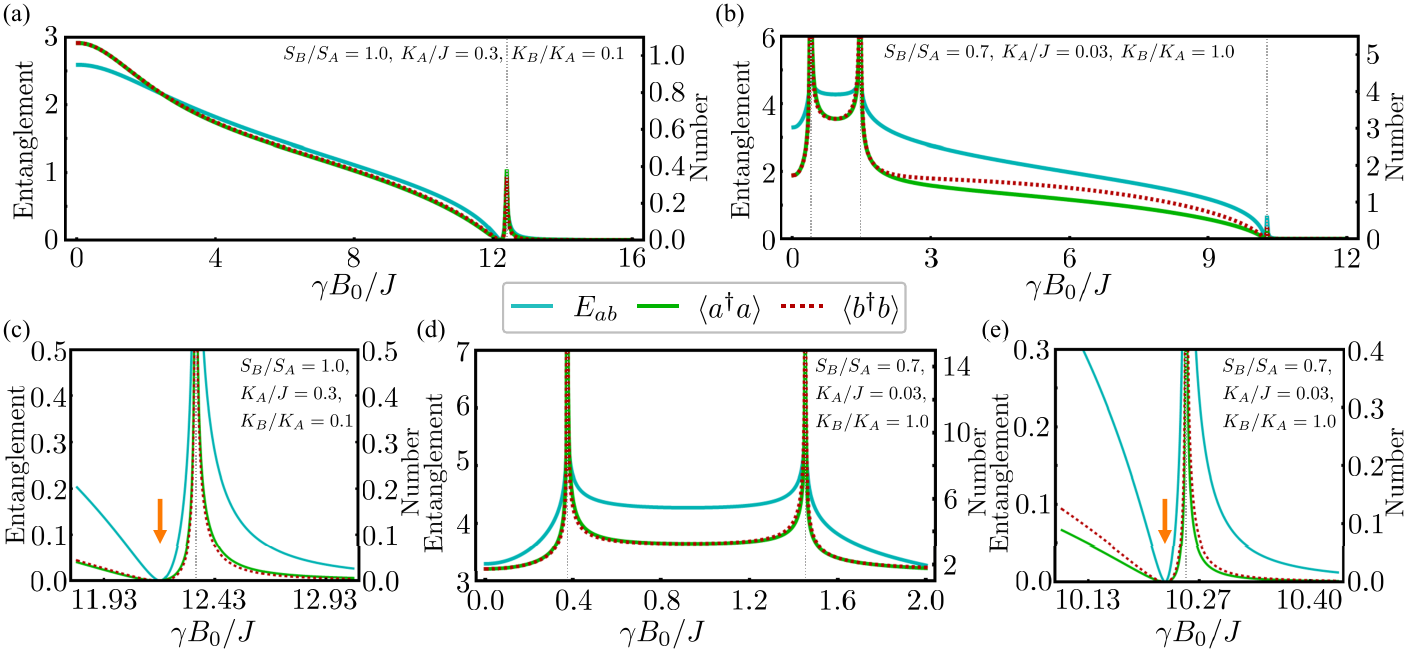}
    \caption{Magnon number and local entanglement entropy. In all panels, the magnon numbers for the magnon components, $\langle a^{\dagger}a\rangle$ and $\langle b^{\dagger}b\rangle$, and the local von Neumann entropy between the magnons $E_{ab}$ are plotted against the magnetic field strength, following the plot legend in the middle. The two left panels (a) and (c) represent the antiferromagnetic regime, while the three right panels (b), (d), and (e) represent the ferrimagnetic regime. The lower panels are plotted near the critical points. The arrows in (c) and (e) indicate the points where the magnon number and entropy are suppressed.     }
    \label{Fig_Magnon_Entanglement_1}
\end{figure*}

\subsection{Critical entanglement of magnons}
We now expand the spin Hamiltonian into the magnon Hamiltonian near the mean-field solution. Regardless of the spin orderings, the general magnon Hamiltonian density in momentum space is given by,
\begin{equation}
\begin{aligned}
\mathcal{H}^{\text{m}}_{\mathbf{k}} = & 
\frac{\omega^{A}_{0}}{2}a^{\dagger}_{\mathbf{k}}a_{\mathbf{k}}+\frac{\omega^{B}_{0}}{2}b^{\dagger}_{\mathbf{k}}b_{\mathbf{k}}+ \omega^{A}_{s}a_{\mathbf{k}}a_{-\mathbf{k}} + \omega^{B}_{s} b_{\mathbf{k}}b_{-\mathbf{k}}\\
&+h^{-}_{\mathbf{k}}a_{\mathbf{k}}b_{-\mathbf{k}}+h^{+}_{\mathbf{k}}a^{\dagger}_{\mathbf{k}}b_{\mathbf{k}}+\text{H.c.},
\label{Eq_H_mag_1}
\end{aligned}
\end{equation}
where $\omega^{\sigma}_{0}= -JS_{\bar{\sigma}} n_{1}\cos(\theta_{A}-\theta_{B})+K_{\sigma}S_{\sigma} (2-3\sin^{2}\theta_{\sigma})+\gamma B_{0}\sin\theta_{\sigma}$, $\omega^{\sigma}_{s}=(K_{\sigma}S_{\sigma} /2)\sin^{2}\theta_{\sigma}$, and $h^{\pm}_{\mathbf{k}}=(J\sqrt{S_{A}S_{B}}n_{1}\gamma_{\mathbf{k}}/2)(1\pm \cos(\theta_{A}-\theta_{B}))$. Here, $\theta_{\sigma}$ represents the mean-field solution and is a function of system parameters, including the magnetic field strength. We neglected the higher-order magnon terms as these terms are typically suppressed by the size of the magnet; see Appendix~\ref{Appx_Higher}. We note that there are several types of squeezing terms. First, there are squeezing processes among magnons within a single sublattice, whose magnitude is given by $\omega_s^\sigma$. They are single-mode and two-mode squeezing for the Kittel (${\bf k}=0$) mode and finite momentum modes, respectively. In addition, there are two-mode squeezing terms between different sublattice magnons proportional to $h^{-}_{\bf k}$. In addition, there is a beam-splitter-type magnon-hopping term between the sublattice magnons with a hopping strength $h^{+}_{\bf k}$. These single-mode and two-mode squeezing terms, together with the beam-splitter type hopping interaction, offer an important toolbox to engineer magnon states and magnon-magnon coupling in magnetic materials~\cite{shiota2020tunable,sud2020tunable,wang2024ultrastrong}. The coefficients of these squeezing terms as a function of the magnetic field are shown in Fig.~\ref{FIG_CouplingStrength_1}, which show the relative significance of each term. The role of the single-mode squeezing term in modifying the spin angular momentum of the magnon eigenmode~\cite{kamra2016super,kamra2017spin,lee2023cavity} and the criticality of the ferromagnet~\cite{lee2023cavity} has been investigated. As we demonstrate below, the interplay between various inter- and intra-sublattice squeezing processes drives the phase transition among the spin orderings investigated previously, through the critical entanglement of magnons.

We obtain the frequency of the magnon eigenmodes by diagonalizing the magnon Hamiltonian [Eq.~(\ref{Eq_H_mag_1})]. We plot the frequency of the uniform mode $\omega=\omega_{\mathbf{k}=0}$ across the magnetic field strength $B_{0}$ for both the antiferromagnetic and ferrimagnetic regimes in Fig.~\ref{Fig_Ordering_Magnon_1}(a) and (b), respectively. Notably, we observe the closure of the frequency gap at all boundaries of the phases examined at the mean-field level, signifying that the softening of the magnon modes results in the magnetic instabilities and the phase transitions. The magnon gap closes at a single critical point in both the antiferromagnetic regime and the ferrimagnetic regime with small spin asymmetry [Fig.~\ref{Fig_Ordering_Magnon_1}(a)]. Meanwhile, it closes at three critical field strengths in the ferrimagnetic regime with larger spin asymmetry [Fig.~\ref{Fig_Ordering_Magnon_1}(b)]. The gap closure follows a square root behavior, $\omega \propto |B_{0} - B^{c}_{0}|^{1/2}$, characteristic of a mean-field second-order phase transitions, where $B^{c}_{0}$ denotes the critical magnetic field strength. We also plot the frequency band $\omega_{\mathbf{k}}$ on the $k_{x}$-$k_{y}$ plane for fixed $B_{0}$ and $k_{z}$ in Fig.~\ref{Fig_Ordering_Magnon_1}(e) and (f). In the absence of a magnetic field, the two bands are separated by an energy scale proportional to the anisotropy strength and the spin angular momentum differences. As the magnetic field strength increases, the upper band rises and becomes nearly flat at a specific field strength; see the middle panels in Fig.~\ref{Fig_Ordering_Magnon_1}(e) and (f). Notably, the upper band becomes almost flat in the ferrimagnetic regime~\cite{wang2024broad}. The flatness depends on parameters such as the spin angular momentum ratio. Beyond this point, its curvature becomes conversed under a stronger magnetic field.

Let us now investigate the correlation between the different sublattice magnons by calculating the entanglement. The various squeezing terms and hopping terms in the magnon Hamiltonian [Eq.~(\ref{Eq_H_mag_1})] contribute to the entanglement, unlike the case where the two-mode squeezing parameter entirely represented the entanglement between sublattices~\cite{kamra2019antiferromagnetic,wuhrer2022theory}. Here, we focus on the entanglement property of the uniform modes, which becomes critical, and we present the analysis of non-uniform modes with finite momentum in Appendix~\ref{Appx_FiniteMomentum}. To measure the entanglement, we calculate and plot the von Neumann entropy $E_{ab}$ between the uniform magnon modes $a=a_{\mathbf{k}=0}$ and $b=b_{\mathbf{k}=0}$ of different sublattices in Fig.~\ref{Fig_Magnon_Entanglement_1} using the covariance matrix~\cite{pirandola2009correlation}.  The entanglement is nonzero in the absence of the magnetic field due to two-mode squeezing from antiferromagnetic exchange interaction. As the magnetic field increases, the entanglement diverges at all critical points, indicating the divergence of the correlation, similar to the divergence of single-mode squeezing in the easy-axis ferromagnet~\cite{lee2023cavity}. Note that diverging entanglement appears finite in the figure due to numerical precision. It is interesting to observe that the two-mode squeezing terms proportional to $h^{-}_{\bf k}$ become zero at the critical point; therefore, the divergence of the entanglement is due to the interplay between the single-mode squeezing and the number-conserving terms. Our analysis shows that the anisotropy and the perpendicular magnetic field can induce arbitrarily large entanglement entropy of magnons, facilitating the engineering of magnon-magnon entanglement~\cite{zou2020tuning,yuan2020enhancement,shim2022enhanced}.

We also plot the number of magnons $\langle a^{\dagger}a\rangle$ and $\langle b^{\dagger}b\rangle$ in the ground state as a function of the magnetic field strength in Fig.~\ref{Fig_Magnon_Entanglement_1}. These values are nonzero across all ranges due to the squeezing. Unlike the entropy, the magnon numbers are distinct for each component, with the difference proportional to the asymmetry in anisotropy strength and spin amplitude. The magnon numbers also diverge at all critical points as the infinite bipartite entropy between two bosonic modes requires an infinite fluctuation in the number of excitations. We also find an interesting non-monotonic behavior of the magnon number; namely, the magnon number is strongly suppressed before this divergence at the phase transition (see arrows in Fig.~\ref{Fig_Magnon_Entanglement_1}(c) and (e)). At this point, the coefficients of the squeezing terms $\omega^{\sigma}_{s}$ and $h^{-}_{\bf k}$ become nearly identical. As the magnetic field increases, both angles $\theta_{\sigma}$ approach $\pi/2$, leading to a situation where $\omega^{\sigma}_{s}$ and $h^{-}_{\bf k}$ become vanishingly small, but remain finite if $J \gg K_{\sigma}$. As a result, the coefficients of all squeezing terms have nearly identical small amplitudes, leading to the highly suppressed magnon number in the ground state. Despite this suppression, the ground state never becomes an exact vacuum state under any parameter conditions. Both the magnon number suppression and divergence are experimentally observable consequences of the competing magnon squeezing effects.

\section{Cavity magnonics} \label{Sec_CavityMagnonics}
A magnet in a cavity, known as the cavity magnonics, realizes a strong magnon-photon coupling from the interaction between the spin and the magnetic cavity field~\cite{soykal2010strong,yuan2022quantum}. The cavity magnonics has opened opportunities for the precision measurement of the magnon~\cite{lachance2020entanglement}, the non-Hermitian physics~\cite{harder2018level,zhang2019higher,wang2019nonreciprocity}, the quantum transducer~\cite{hisatomi2016bidirectional}, and the quantum state preparation~\cite{bittencourt2019magnon}. A previous study suggested the measurement of the single-mode magnon squeezing in a cavity magnonics setup with the easy-axis ferromagnet by observing a cavity photon population~\cite{lee2023cavity}. It has been demonstrated that the magnon-photon coupling translates the instability of the magnet into photon condensation in the cavity, inducing the superradiant phase transition of the cavity magnonics system. Motivated by this idea, we explore the possibility of observing the superradiant phase transition driven by the magnetic phase transition of anti-ferromagnets and ferrimagnets. This approach offers an experimentally feasible way to observe the magnetic phase transition through cavity photons.

We consider an antiferromagnet or a ferrimagnet, described by Eq.~(\ref{Eq_General_Ham}), placed within a cavity where the cavity magnetic field is oriented along the $z$ direction. As we shall see, this choice of the cavity field direction ensures that the magnetic instabilities induce the critical phenomena of the cavity such as the photon condensation and diverging photon fluctuation. For other choices of the cavity field directions, we present their analysis in Appendix~\ref{Appx_AlternativeDirection}. We focus solely on the uniform magnon mode, which couples to the uniform cavity field, assuming a small magnet size relative to the wavelength of the cavity field to ensure the validity of the dipole approximation. Additionally, for simplicity, we assume symmetric bare spin-cavity coupling strengths for both sublattices. The Hamiltonian for the combined system of the cavity and the magnet is given by
\begin{equation}
    H'_{0} = H_{0}+ \omega_{c}c^{\dagger}c - g (c+c^{\dagger})\sum_{i,\sigma} \sqrt{\frac{2}{S_{\sigma} N}} S^{z}_{\sigma,\mathbf{r}_{i}},
\end{equation}
where $H_{0}$ is the spin Hamiltonian in Eq.~(\ref{Eq_General_Ham}), $c$ is the cavity photon annihilation operator, $\omega_{c}$ is the cavity frequency, $N$ is the number of spins, and $g$ is the bare spin-cavity coupling strength. Introducing the cavity photon condensation $\langle c\rangle = c_{0} $, the mean-field energy per spin is expressed as follows.
\begin{equation}
\begin{aligned}
    \mathcal{E}_{\rm cm} =& \mathcal{E}_{\rm m}-2g c_{0}\sum_{\sigma} \sqrt{\frac{2S_{\sigma}}{N}} \cos\theta_{\sigma}+\frac{\omega_{c}}{N} c^{2}_{0},
\end{aligned}
\label{Eq_MFE_CM_1}
\end{equation}
where $\mathcal{E}_{\rm m}$ is the mean-field energy of the spin without the cavity in Eq.~(\ref{Eq_MF_1}).

\begin{figure}[t!]
    \centering
    \includegraphics[width=\linewidth]{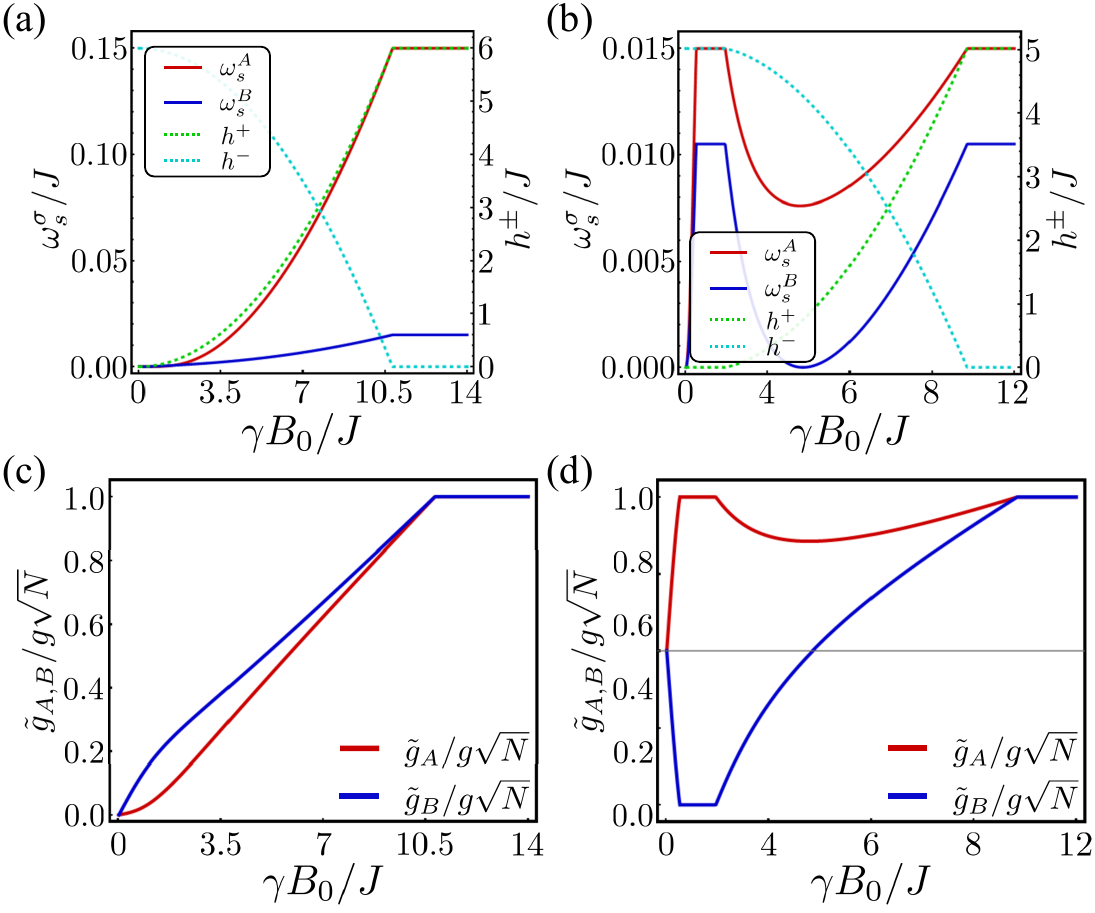}
    \caption{  Coefficients of the magnon Hamiltonian [Eq.~(\ref{Eq_H_mag_1})] as a function of the magnetic field strength $B_0$ in (a) the antiferromagnetic and (b) the ferrimagnetic regimes. Here, $h^{\pm}=h^{\pm}_{\mathbf{k}=0}$. Effective coupling strengths $\tilde{g}_{A,B}$ of sublattices $A$ and $B$ as a function of the magnetic field strength $B_0$ in (c) the antiferromagnetic and (d) the ferrimagnetic regimes. }  
    \label{FIG_CouplingStrength_1}
\end{figure}

\begin{figure*}[t!]
    \includegraphics[width=\linewidth]{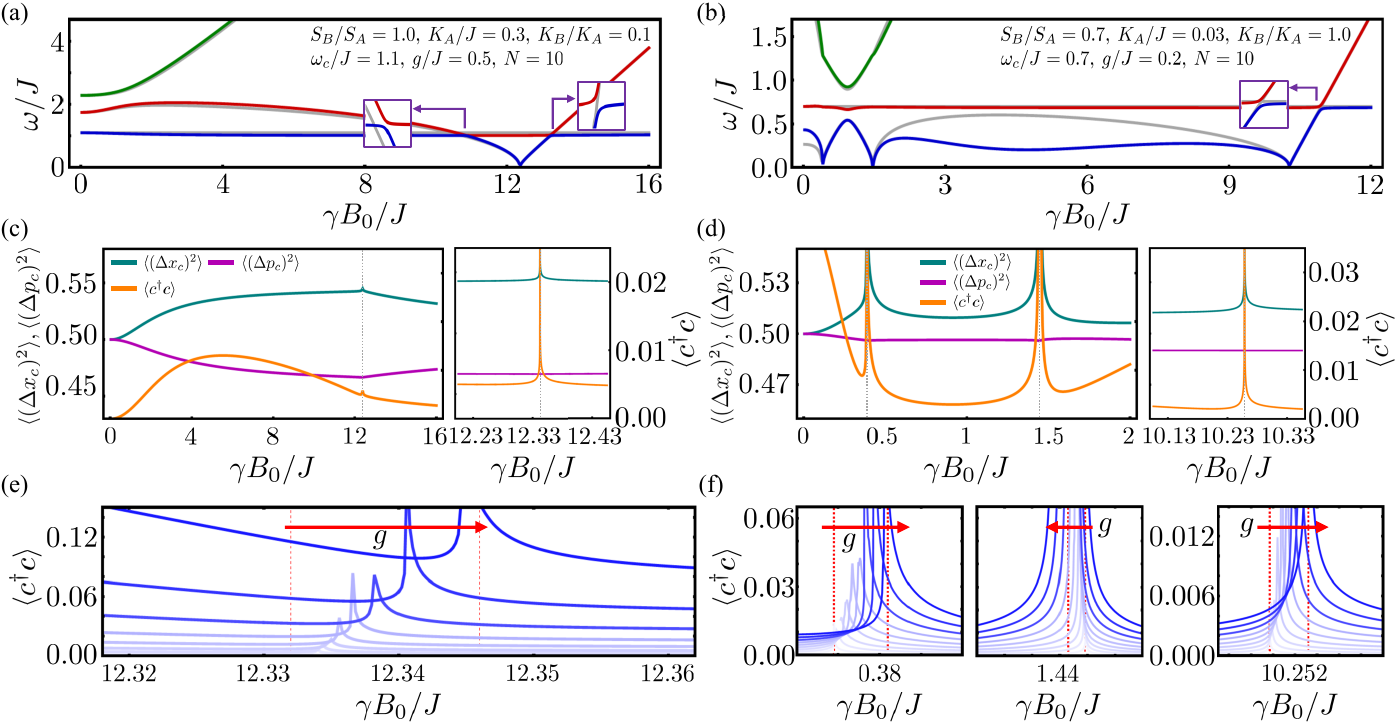}
    \caption{Magnon-polariton states in the cavity magnonics system. The left three panels pertain to the antiferromagnetic regime, while the right three panels correspond to the ferrimagnetic regime. (a, b) The frequency spectrum of the magnon-polariton, as a function of magnetic field strength, with gray lines indicating the frequency levels in the absence of magnon-photon interaction. (c, d) Quadratures of the photon canonical operators and photon number in the ground state across varying magnetic field strengths near the critical point. (e, f) Photon numbers for different bare spin-cavity coupling strengths along the magnetic field strength, with red arrows indicating the shifts in the graphs as the coupling strength increases. }
    \label{Fig_Cavity_Magnonics_1}
\end{figure*}

Quantum fluctuation of both cavity field and magnons around the mean-field solution is governed by the cavity magnonics Hamiltonian,
\begin{equation}
\begin{aligned}
H^{\text{cm}}=& \frac{ \omega^{A}_{m} }{2}a^{\dagger}a+\frac{ \omega^{B}_{m} }{2}b^{\dagger}b+\frac{\omega_{c}}{2}\tilde{c}^{\dagger}\tilde{c}+ \omega^{A}_{s}a^{2} \\
&+ \omega^{B}_{s} b^{2} +h^{-}ab+h^{+}a^{\dagger}b-i\tilde{g}_{A}(a\tilde{c}+a\tilde{c}^{\dagger})\\
&-i\tilde{g}_{B}(b\tilde{c}+b\tilde{c}^{\dagger})+\text{H.c.},
\label{Eq_H_CM_1}
\end{aligned}
\end{equation}
where 
\begin{equation}
    \omega^{\sigma}_{m} = \omega^{\sigma}_{0} + \frac{4g^{2}}{\omega_{c}\sqrt{S_{\sigma}}}\cos\theta_{\sigma}\sum_{\sigma'}\sqrt{S_{\sigma'}} \cos\theta_{\sigma'}
\end{equation}
is the renormalized magnon frequency, $\tilde{c}=c-c_{0}$ is the effective cavity photon operator with the photon condensation
\begin{equation}
c_{0} = \frac{g\sqrt{2N}}{\omega_{c}} (\sqrt{S_{A}}\cos\theta_{A}+\sqrt{S_{B}}\cos\theta_{B}),
\end{equation}
$a=a_{\mathbf{k}=0}$ and $b=b_{\mathbf{k}=0}$ are magnon operators of uniform modes, $h^{\pm}=h^{\pm}_{\mathbf{k}=0}$, and $\tilde{g}_{\sigma}=g\sin\theta_{\sigma}$. The effective magnon-photon coupling strength $\tilde{g}_{\sigma}$ depends on the magnetic field, reaching its maximum amplitude in the normal phase where the spins are aligned either parallel or antiparallel to each other (Fig.~\ref{FIG_CouplingStrength_1}). We also note that photon condensation occurs with non-zero order parameter $c_0$ only when the net angular momentum exists (green lines in Fig.~\ref{Fig_Ordering_Magnon_1}), indicating the emergence of the superradiant phase. We diagonalize the cavity magnonics Hamiltonian given in Eq.~(\ref{Eq_H_CM_1}), which leads to the hybridization of cavity photons and magnons into a magnon-polariton.
We plot the frequency dispersion of the magnon-polariton for both antiferromagnetic and ferrimagnetic regimes in Fig.~\ref{Fig_Cavity_Magnonics_1}(a) and (b). The avoided crossings in the dispersion reveal the magnon-photon interaction despite their small size in the plot. Notably, the low-lying magnon-polariton softens and becomes critical at the superradiant phase transition point with a power-law with an exponent $1/2$, $\omega \propto |B_{0}-B^{c}_{0}|^{1/2}$. We also investigate the quantum fluctuation of the cavity field by calculating the variance of quadrature variables defined as follows,
\begin{equation}
    x_{c} = \frac{1}{\sqrt{2}}(c+c^{\dagger}),\quad p_{c} = \frac{1}{\sqrt{2}i}(c-c^{\dagger}),
\label{Eq_CavPh1}
\end{equation}
which can be observed using cavity photon measurements. We plot the variance of quadratures $\langle (\Delta x_{c})^{2}\rangle$ and $\langle (\Delta p_{c})^{2}\rangle$, along with the number of cavity photons $\langle c^{\dagger}c\rangle$ in Figs.~\ref{Fig_Cavity_Magnonics_1}(c) and (d). Both $\langle (\Delta x_{c})^{2}\rangle$ and $\langle c^{\dagger}c\rangle$ diverge at all critical points with a power-law with an exponent $-1/2$, an important characteristic of a superradiant phase transition~\cite{Lambert2004,hwang2015}. We note that the variance of $p_{c}$, while being below the minimum uncertainty value of $1/2$ due to squeezing, remains finite. Consequently, the product of variances, $\langle (\Delta x_{c})^{2} \rangle \langle (\Delta p_{c})^{2} \rangle$, which quantifies the uncertainty of the cavity field, diverges at the critical point with the same exponent. Furthermore, the critical field strengths of all critical points are influenced by the magnon-photon coupling. As illustrated in Figs.~\ref{Fig_Cavity_Magnonics_1}(e) and (f), the field strength at which the photon number diverges shifts upwards as the coupling strength increases.

The closing energy gap of the magnon-polariton and the divergences of fluctuations demonstrate that the magnetic phase transitions in the magnet indeed induce critical phenomena of the cavity magnonics system. The spontaneous breaking of the mirror symmetry in the antiferromagnets leads to the symmetry-broken phase of the combined system of the cavity and the magnets, characterized by photon condensation. In particular, repeated superradiant phase transitions back and forth between the normal and superradiant phase occur for the cavity magnonics system with ferrimagnets due to the consecutive magnetic phase transitions we discovered in the previous sections. Therefore, our analysis demonstrates that one could observe the magnetic phase transition with the critical magnon entanglement through the superradiant phase transitions in the cavity magnonics system with either the antiferromagnet or the ferrimagnet by measuring the cavity photon. Moreover, it offers a unique opportunity to realize repeated superradiant phase transitions by modulating a single control parameter, showcasing a phenomenon not commonly observed in quantum optical systems.

\section{Conclusion} \label{Sec_Conclusion}
We investigated the antiferromagnetic magnon state with non-collinear spin ordering, where the easy-axis anisotropy and the external magnetic field are perpendicular to each other. As the magnetic field increases, several critical points emerge, associated with the mirror symmetry of the spin ordering. At these critical points, magnon entanglement between the two sublattices and the magnon number exhibit divergence, characteristic of critical phenomena. Introducing a cavity and magnon-photon coupling transforms these magnetic instability phase transitions into superradiant phase transitions of the cavity magnonics system, leading to photon condensation. At the critical point, the cavity photon is strongly squeezed, and its number diverges. We proposed using the cavity photon measurement as a method to observe the magnetic phase transition and the criticality of the magnon.

\acknowledgements
J.M.L. and H.-W.L. acknowledge support from the National Research Foundation of Korea (NRF) grant funded by the Korean government (MSIT) (No. RS-2024-00410027). M.-J.H. was supported by the Startup Fund from Duke Kunshan University, and Innovation Program for Quantum Science and Technology 2021ZD0301602.

\appendix

\section{Higher-Order Magnon Terms} \label{Appx_Higher}
In the previous sections, we considered only the quadratic terms in the magnon Hamiltonian, neglecting higher-order terms under the assumption that they are effectively suppressed by the size of the magnet. However, near the critical point, where the magnon number increases significantly, it is important to examine the role of these higher-order terms, particularly in the diverging entanglement. To investigate this, we extend the Holstein-Primakoff transformation to include the next-leading-order terms, as described below.
\begin{equation}
\begin{aligned}
S^{+}_{\sigma, \mathbf{r}} &\simeq \sqrt{2S_{\sigma}} m_{\sigma,\mathbf{r}} - \frac{1}{2\sqrt{2S_{\sigma}}} m^{\dagger}_{\sigma,\mathbf{r}}m^{2}_{\sigma,\mathbf{r}},\\
S^{-}_{\sigma, \mathbf{r}} &\simeq \sqrt{2S_{\sigma}} m^{\dagger}_{\sigma,\mathbf{r}} - \frac{1}{2\sqrt{2S_{\sigma}}} m^{\dagger 2}_{\sigma,\mathbf{r}}m_{\sigma,\mathbf{r}},\\
S^{z}_{\sigma, \mathbf{r}} &\simeq S_{\sigma}- m^{\dagger}_{\sigma,\mathbf{r}}m_{\sigma,\mathbf{r}},
\end{aligned}
\end{equation}
where $\sigma=A,B$ denotes the sublattice, $m_{A,\mathbf{r}}=a_{\bf r}$ and $m_{B,\mathbf{r}}=b_{\bf r}$ represent the magnon annihilation operators.

\begin{figure}[t!]
    \centering
    \includegraphics[width=0.98\linewidth]{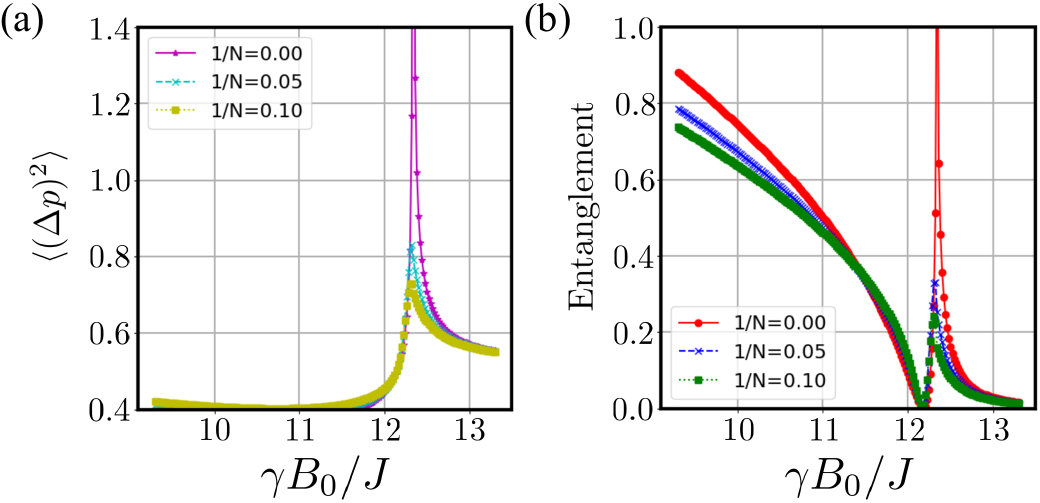}
    \caption{  (a) The quadrature of the magnon along the $p$ direction and (b) the magnon-magnon entanglement between sublattices $A$ and $B$ are plotted along the magnetic field strength $B_0$ near the critical point. The insets denote the strength of the higher-order contribution of the magnon. }  
    \label{FIG_HigherOrder_1}
\end{figure}

Using the extended Holstein-Primakoff transformation outlined above, we expanded the spin Hamiltonian [Eq.~(\ref{Eq_General_Ham})], which includes the exchange interaction, easy-axis anisotropy, and Zeeman interaction. This expansion introduces third-order terms involving three magnon operators as the next-leading-order contributions as follows,
\begin{equation}
H^{(3)}_{\rm h}= H^{(3)}_{J}+H^{(3)}_{K}+H^{(3)}_{B},
\end{equation}
where
\begin{equation}
\begin{aligned}
    H^{(3)}_{J} &= i\frac{Jn_{1}}{\sqrt{N}}\text{cssc}_{\theta}\Big( 
    \sqrt{\frac{S_{A}}{2}}ab^{\dagger}b -\sqrt{\frac{S_{B}}{2}}a^{\dagger}ab\\
    &+\frac{S_{B}}{4\sqrt{2S_{A}}}a^{\dagger}a^{2}- \frac{S_{A}}{4\sqrt{2S_{B}}}b^{\dagger}b^{2}-\text{H.c.} \Big),
\end{aligned}
\end{equation}
\begin{equation}
\begin{aligned}
    H^{(3)}_{K} = \frac{i}{\sqrt{N}}\sum_{\sigma}& K_{\sigma}\sin\theta_{\sigma}\cos\theta_{\sigma}\sqrt{\frac{S_{\sigma}}{2}}\Big(m_{\sigma}m^{\dagger}_{\sigma}m_{\sigma}\\
    &+\frac{3}{2}m^{\dagger}_{\sigma}m^{2}_{\sigma}-\text{H.c.} \Big) ,
\end{aligned}
\end{equation}
\begin{equation}
\begin{aligned}
    H^{(3)}_{B} = -\frac{i\gamma B_{0}}{\sqrt{N}}\sum_{\sigma}& \frac{\cos\theta_{\sigma}}{4\sqrt{2S_{\sigma}}} (m^{\dagger}_{\sigma}m^{2}_{\sigma}-\text{H.c.}),
\end{aligned}
\end{equation}
$\text{cssc}_{\theta}=\cos\theta_{A}\sin\theta_{B}-\sin\theta\cos\theta$, $\sigma=A,B$ denotes the sublattice, $m_{A}=a$ and $m_{B}=b$ represent the annihilation operators of the uniform magnon mode. For simplicity, we focus on the uniform mode although the higher-order terms generally include non-uniform modes with finite momentum. In the normal phase, these third-order terms vanish due to the mirror symmetry so we additionally calculate the fourth-order terms which are next-leading-order contributions in the normal phase as follows,
\begin{equation}
H^{(4)}_{\rm h}= H^{(4)}_{J}+H^{(4)}_{K},
\end{equation}
where
\begin{equation}
\begin{aligned}
    H^{(4)}_{J} &= \frac{Jn_{1}}{8N}\sqrt{\frac{S_{A}}{S_{B}}} (c^{-}_{\theta} a-c^{+}_{\theta}a^{\dagger} )b^{\dagger}b^{2}\\
    &+\frac{Jn_{1}}{8N} \sqrt{\frac{S_{B}}{S_{A}}}a^{\dagger}a^{2}(c^{-}_{\theta} b-c^{+}_{\theta}b^{\dagger} ) \\
    &+\frac{Jn_{1}}{2N}\sin\theta_{A}\sin\theta_{B}a^{\dagger}ab^{\dagger}b +\text{H.c.},
\end{aligned}
\end{equation}
\begin{equation}
\begin{aligned}
    H^{(4)}_{K} = \frac{1}{8N}\sum_{\sigma}& K_{\sigma}\sin^{2}\theta_{\sigma} (m_{\sigma}-m^{\dagger}_{\sigma})\\
    &\times (m^{\dagger}_{\sigma}m^{2}_{\sigma}-m^{\dagger 2}_{\sigma}m_{\sigma})+\text{H.c.},
\end{aligned}
\end{equation}
and $c^{\pm}_{\theta}=\cos(\theta_{A}-\theta_{B})\pm 1$.

We investigate the contribution of higher-order terms in $H^{(3)}_{\rm h}+H^{(4)}_{\rm h}$ to the critical behavior by numerically computing the quadrature variance along the $p$ direction and the magnon-magnon entanglement near the critical point using the QuTiP package~\cite{johansson2012qutip}. In the antiferromagnetic regime, we treat the factor $1/N$ as a control parameter that modulates the strength of the quartic term. For the magnon at sublattice $A$, we define the quadrature variable $p = -i(a - a^{\dagger})/\sqrt{2}$ and calculate its variance $\langle (\Delta p)^{2} \rangle = \langle (p-\langle p\rangle)^{2} \rangle$. The results are shown in Fig.~\ref{FIG_HigherOrder_1}. We observe that the divergence of both quantities is mitigated upon the introduction of the higher-order terms, which instead leads to an enhancement of these quantities, with a sharp peak at the critical point. Thus, we conclude that the higher-order terms regularize the diverging critical behavior; however, further investigation, including the consideration of finite momentum, is required for a more precise understanding.

\begin{figure}[t]
    \centering
    \includegraphics[width=\linewidth]{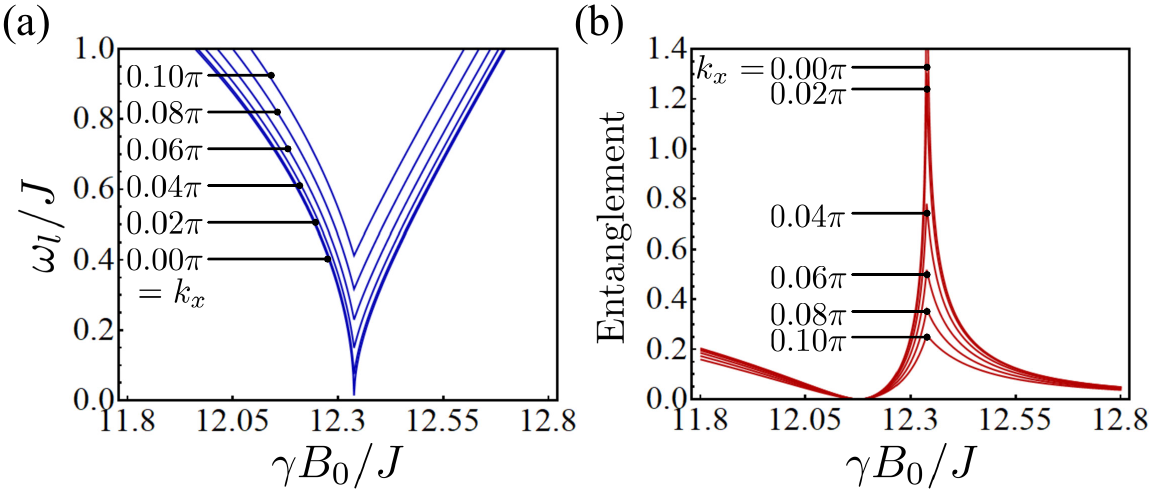}
    \caption{  (a) The frequency of the magnon $\omega_{l}$ corresponding to the lower dispersion branch and (b) the magnon-magnon entanglement between sublattices $A$ and $B$ are plotted for the antiferromagnetic regime along the magnetic field strength $B_0$. Here, $k_y$ and $k_z$ are fixed to zero, reducing the analysis to variations along $k_x$.}  
    \label{FIG_Finite_1}
\end{figure}

\section{Magnon Modes with Finite Momentum} \label{Appx_FiniteMomentum}
In Sec.~\ref{Sec_OrderingsMagnons}, we discussed the magnon-magnon entanglement between sublattices, focusing exclusively on the uniform magnon modes at the $\Gamma$ point. This focus is justified by the fact that these uniform modes drive the magnetic phase transition by closing their frequency gap at the critical magnetic field condition. However, it is natural to ask about the entanglement properties of other non-uniform modes with finite momentum near the critical point.

To explore this, we calculate the magnon frequency and magnon-magnon entanglement for finite momentum $\mathbf{k}$ in Eq.~(\ref{Eq_H_mag_1}). For simplicity, we consider the antiferromagnetic regime and restrict the finite momentum to the $x$ direction. The results, shown in Fig.~\ref{FIG_Finite_1}, reveal that for non-uniform modes, the frequency gap is reduced but does not close at the critical point. Similarly, the entanglement is enhanced near the critical point but does not diverge to infinity. Despite this, the sharp reduction in the frequency gap and the pronounced enhancement of entanglement make the phase transition observable even in the non-uniform modes.

\begin{figure}[t!]
    \centering
    \includegraphics[width=\linewidth]{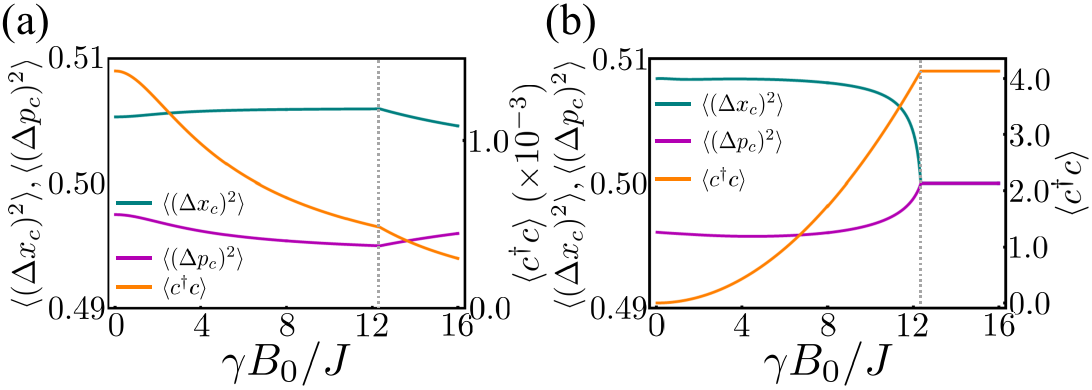}
    \caption{  Quadratures of the photon canonical operators and photon number in the antiferromagnetic regime for cavity field orientations along (a) the $x$ direction and (b) the $y$ direction.  }  
    \label{FIG_Alternative_1}
\end{figure}

\section{Cavity Fields in Alternative Directions} \label{Appx_AlternativeDirection}
In Sec.~\ref{Sec_CavityMagnonics}, we assumed the cavity field was aligned along the $z$ axis, parallel to the easy-axis anisotropy. However, in general, the cavity field can be oriented in arbitrary directions. Here, we discuss cases where the cavity field is aligned along the $x$ axis or the $y$ axis. For both orientations, the cavity-spin Hamiltonian takes the following form,
\begin{equation}
    H'_{0} = H_{0}+ \omega_{c}c^{\dagger}c - g (c+c^{\dagger})\sum_{i,\sigma} \sqrt{\frac{2}{S_{\sigma} N}} S^{x\: \text{or}\: y}_{\sigma,\mathbf{r}_{i}},
\end{equation}
where $H_{0}$ is the spin Hamiltonian defined in Eq.~(\ref{Eq_General_Ham}), and other parameters are consistent with the previous sections.

Regardless of the cavity field direction, one can generally expect instability in the Hamiltonian due to the spin-cavity coupling. In particular, when the cavity field is aligned along the $x$ axis, additional instability may arise, potentially causing the spins to rotate toward the $x$ axis if the coupling is sufficiently strong. However, we do not consider these additional instabilities here. Instead, we focus on verifying whether the magnetic instability of the spin Hamiltonian, as studied in previous sections, propagates to the criticality of the cavity photon, leading to photon condensation and diverging cavity fluctuation.

By performing a mean-field analysis, one can derive the cavity magnonics Hamiltonian based on the classical spin configuration and photon condensation. The photon condensation is given by,
\begin{equation}
    c_{0} = 0\: \text{or}\: \frac{g}{\omega_{c}} \sum_{\sigma} \sqrt{2NS_{\sigma}} \sin\theta_{\sigma},
\end{equation}
depending on the cavity field orientation. Notably, photon condensation does not occur in the $x$-axis case. The total cavity magnonics Hamiltonian is expressed as
\begin{equation}
\begin{aligned}
H^{\rm cm}_{x\: \text{or}\:y} = H^{\rm m}_{\mathbf{k}=0} + \omega_{c}\tilde{c}^{\dagger}\tilde{c} + H'_{\rm int},
\end{aligned}
\end{equation}
where the cavity-spin interaction term is given by,
\begin{equation}
\begin{aligned}
    H'_{\rm int} &= -g(\tilde{c}+\tilde{c}^{\dagger})\sum_{\sigma}(m_{\sigma}+m^{\dagger}_{\sigma}),
\end{aligned}
\end{equation}
or
\begin{equation}
\begin{aligned}
   H'_{\rm int} =& ig(\tilde{c}+\tilde{c}^{\dagger})\sum_{\sigma}\cos\theta_{\sigma}(m_{\sigma}-m^{\dagger}_{\sigma})\\
   &+2gc_{0}\sum_{\sigma}\sqrt{\frac{2}{S_{\sigma}N}}\sin\theta_{\sigma}m^{\dagger}_{\sigma}m_{\sigma},
\end{aligned}
\end{equation}
depending on the cavity field orientation. Here, $\sigma=A,B$ denotes the sublattice, and $m_{A}=a$ and $m_{B}=b$ represent the magnon annihilation operators. The magnon-photon coupling strength remains invariant under all parameters of the spin Hamiltonian in the case of the $x$ axis. For the $y$-axis case, the magnon-photon coupling is absent in the normal phase, although photon condensation causes a shift in the magnon frequency.

For simplicity, we consider the antiferromagnetic regime. We diagonalize the cavity magnonics Hamiltonian and present the photon number and the quadratures of the photon canonical operators in Fig.~\ref{FIG_Alternative_1}. Unlike the case where the cavity field is aligned along the $z$ axis, these quantities do not diverge at the critical point where the magnon-polariton frequency vanishes. Therefore, we conclude that aligning the cavity field along the $z$ axis is essential for observing the magnetic phase transition through cavity photon measurements.

\bibliography{BibRef}

\begin{thebibliography}{50}%
\makeatletter
\providecommand \@ifxundefined [1]{%
 \@ifx{#1\undefined}
}%
\providecommand \@ifnum [1]{%
 \ifnum #1\expandafter \@firstoftwo
 \else \expandafter \@secondoftwo
 \fi
}%
\providecommand \@ifx [1]{%
 \ifx #1\expandafter \@firstoftwo
 \else \expandafter \@secondoftwo
 \fi
}%
\providecommand \natexlab [1]{#1}%
\providecommand \enquote  [1]{``#1''}%
\providecommand \bibnamefont  [1]{#1}%
\providecommand \bibfnamefont [1]{#1}%
\providecommand \citenamefont [1]{#1}%
\providecommand \href@noop [0]{\@secondoftwo}%
\providecommand \href [0]{\begingroup \@sanitize@url \@href}%
\providecommand \@href[1]{\@@startlink{#1}\@@href}%
\providecommand \@@href[1]{\endgroup#1\@@endlink}%
\providecommand \@sanitize@url [0]{\catcode `\\12\catcode `\$12\catcode `\&12\catcode `\#12\catcode `\^12\catcode `\_12\catcode `\%12\relax}%
\providecommand \@@startlink[1]{}%
\providecommand \@@endlink[0]{}%
\providecommand \url  [0]{\begingroup\@sanitize@url \@url }%
\providecommand \@url [1]{\endgroup\@href {#1}{\urlprefix }}%
\providecommand \urlprefix  [0]{URL }%
\providecommand \Eprint [0]{\href }%
\providecommand \doibase [0]{https://doi.org/}%
\providecommand \selectlanguage [0]{\@gobble}%
\providecommand \bibinfo  [0]{\@secondoftwo}%
\providecommand \bibfield  [0]{\@secondoftwo}%
\providecommand \translation [1]{[#1]}%
\providecommand \BibitemOpen [0]{}%
\providecommand \bibitemStop [0]{}%
\providecommand \bibitemNoStop [0]{.\EOS\space}%
\providecommand \EOS [0]{\spacefactor3000\relax}%
\providecommand \BibitemShut  [1]{\csname bibitem#1\endcsname}%
\let\auto@bib@innerbib\@empty
\bibitem [{\citenamefont {Kittel}\ \emph {et~al.}(1996)\citenamefont {Kittel}, \citenamefont {McEuen},\ and\ \citenamefont {McEuen}}]{kittel1996introduction}%
  \BibitemOpen
  \bibfield  {author} {\bibinfo {author} {\bibfnamefont {C.}~\bibnamefont {Kittel}}, \bibinfo {author} {\bibfnamefont {P.}~\bibnamefont {McEuen}},\ and\ \bibinfo {author} {\bibfnamefont {P.}~\bibnamefont {McEuen}},\ }\href@noop {} {\emph {\bibinfo {title} {Introduction to solid state physics}}},\ Vol.~\bibinfo {volume} {8}\ (\bibinfo  {publisher} {Wiley New York},\ \bibinfo {year} {1996})\BibitemShut {NoStop}%
\bibitem [{\citenamefont {Jungwirth}\ \emph {et~al.}(2016)\citenamefont {Jungwirth}, \citenamefont {Marti}, \citenamefont {Wadley},\ and\ \citenamefont {Wunderlich}}]{jungwirth2016antiferromagnetic}%
  \BibitemOpen
  \bibfield  {author} {\bibinfo {author} {\bibfnamefont {T.}~\bibnamefont {Jungwirth}}, \bibinfo {author} {\bibfnamefont {X.}~\bibnamefont {Marti}}, \bibinfo {author} {\bibfnamefont {P.}~\bibnamefont {Wadley}},\ and\ \bibinfo {author} {\bibfnamefont {J.}~\bibnamefont {Wunderlich}},\ }\bibfield  {title} {\bibinfo {title} {Antiferromagnetic spintronics},\ }\href {https://doi.org/10.1038/nnano.2016.18} {\bibfield  {journal} {\bibinfo  {journal} {Nature nanotechnology}\ }\textbf {\bibinfo {volume} {11}},\ \bibinfo {pages} {231} (\bibinfo {year} {2016})}\BibitemShut {NoStop}%
\bibitem [{\citenamefont {Nakata}\ \emph {et~al.}(2017)\citenamefont {Nakata}, \citenamefont {Simon},\ and\ \citenamefont {Loss}}]{nakata2017spin}%
  \BibitemOpen
  \bibfield  {author} {\bibinfo {author} {\bibfnamefont {K.}~\bibnamefont {Nakata}}, \bibinfo {author} {\bibfnamefont {P.}~\bibnamefont {Simon}},\ and\ \bibinfo {author} {\bibfnamefont {D.}~\bibnamefont {Loss}},\ }\bibfield  {title} {\bibinfo {title} {Spin currents and magnon dynamics in insulating magnets},\ }\href {https://doi.org/10.1088/1361-6463/aa5b09} {\bibfield  {journal} {\bibinfo  {journal} {Journal of Physics D: Applied Physics}\ }\textbf {\bibinfo {volume} {50}},\ \bibinfo {pages} {114004} (\bibinfo {year} {2017})}\BibitemShut {NoStop}%
\bibitem [{\citenamefont {Baltz}\ \emph {et~al.}(2018)\citenamefont {Baltz}, \citenamefont {Manchon}, \citenamefont {Tsoi}, \citenamefont {Moriyama}, \citenamefont {Ono},\ and\ \citenamefont {Tserkovnyak}}]{baltz2018antiferro}%
  \BibitemOpen
  \bibfield  {author} {\bibinfo {author} {\bibfnamefont {V.}~\bibnamefont {Baltz}}, \bibinfo {author} {\bibfnamefont {A.}~\bibnamefont {Manchon}}, \bibinfo {author} {\bibfnamefont {M.}~\bibnamefont {Tsoi}}, \bibinfo {author} {\bibfnamefont {T.}~\bibnamefont {Moriyama}}, \bibinfo {author} {\bibfnamefont {T.}~\bibnamefont {Ono}},\ and\ \bibinfo {author} {\bibfnamefont {Y.}~\bibnamefont {Tserkovnyak}},\ }\bibfield  {title} {\bibinfo {title} {Antiferromagnetic spintronics},\ }\href {https://doi.org/10.1103/RevModPhys.90.015005} {\bibfield  {journal} {\bibinfo  {journal} {Rev. Mod. Phys.}\ }\textbf {\bibinfo {volume} {90}},\ \bibinfo {pages} {015005} (\bibinfo {year} {2018})}\BibitemShut {NoStop}%
\bibitem [{\citenamefont {Kim}\ \emph {et~al.}(2022)\citenamefont {Kim}, \citenamefont {Beach}, \citenamefont {Lee}, \citenamefont {Ono}, \citenamefont {Rasing},\ and\ \citenamefont {Yang}}]{kim2022ferrimagnetic}%
  \BibitemOpen
  \bibfield  {author} {\bibinfo {author} {\bibfnamefont {S.~K.}\ \bibnamefont {Kim}}, \bibinfo {author} {\bibfnamefont {G.~S.}\ \bibnamefont {Beach}}, \bibinfo {author} {\bibfnamefont {K.-J.}\ \bibnamefont {Lee}}, \bibinfo {author} {\bibfnamefont {T.}~\bibnamefont {Ono}}, \bibinfo {author} {\bibfnamefont {T.}~\bibnamefont {Rasing}},\ and\ \bibinfo {author} {\bibfnamefont {H.}~\bibnamefont {Yang}},\ }\bibfield  {title} {\bibinfo {title} {Ferrimagnetic spintronics},\ }\href {https://doi.org/10.1038/s41563-021-01139-4} {\bibfield  {journal} {\bibinfo  {journal} {Nature materials}\ }\textbf {\bibinfo {volume} {21}},\ \bibinfo {pages} {24} (\bibinfo {year} {2022})}\BibitemShut {NoStop}%
\bibitem [{\citenamefont {Jungwirth}\ \emph {et~al.}(2018)\citenamefont {Jungwirth}, \citenamefont {Sinova}, \citenamefont {Manchon}, \citenamefont {Marti}, \citenamefont {Wunderlich},\ and\ \citenamefont {Felser}}]{jungwirth2018multiple}%
  \BibitemOpen
  \bibfield  {author} {\bibinfo {author} {\bibfnamefont {T.}~\bibnamefont {Jungwirth}}, \bibinfo {author} {\bibfnamefont {J.}~\bibnamefont {Sinova}}, \bibinfo {author} {\bibfnamefont {A.}~\bibnamefont {Manchon}}, \bibinfo {author} {\bibfnamefont {X.}~\bibnamefont {Marti}}, \bibinfo {author} {\bibfnamefont {J.}~\bibnamefont {Wunderlich}},\ and\ \bibinfo {author} {\bibfnamefont {C.}~\bibnamefont {Felser}},\ }\bibfield  {title} {\bibinfo {title} {The multiple directions of antiferromagnetic spintronics},\ }\href {https://doi.org/10.1038/s41567-018-0063-6} {\bibfield  {journal} {\bibinfo  {journal} {Nature Physics}\ }\textbf {\bibinfo {volume} {14}},\ \bibinfo {pages} {200} (\bibinfo {year} {2018})}\BibitemShut {NoStop}%
\bibitem [{\citenamefont {Liensberger}\ \emph {et~al.}(2019)\citenamefont {Liensberger}, \citenamefont {Kamra}, \citenamefont {Maier-Flaig}, \citenamefont {Gepr\"ags}, \citenamefont {Erb}, \citenamefont {Goennenwein}, \citenamefont {Gross}, \citenamefont {Belzig}, \citenamefont {Huebl},\ and\ \citenamefont {Weiler}}]{liensberger2019exchange}%
  \BibitemOpen
  \bibfield  {author} {\bibinfo {author} {\bibfnamefont {L.}~\bibnamefont {Liensberger}}, \bibinfo {author} {\bibfnamefont {A.}~\bibnamefont {Kamra}}, \bibinfo {author} {\bibfnamefont {H.}~\bibnamefont {Maier-Flaig}}, \bibinfo {author} {\bibfnamefont {S.}~\bibnamefont {Gepr\"ags}}, \bibinfo {author} {\bibfnamefont {A.}~\bibnamefont {Erb}}, \bibinfo {author} {\bibfnamefont {S.~T.~B.}\ \bibnamefont {Goennenwein}}, \bibinfo {author} {\bibfnamefont {R.}~\bibnamefont {Gross}}, \bibinfo {author} {\bibfnamefont {W.}~\bibnamefont {Belzig}}, \bibinfo {author} {\bibfnamefont {H.}~\bibnamefont {Huebl}},\ and\ \bibinfo {author} {\bibfnamefont {M.}~\bibnamefont {Weiler}},\ }\bibfield  {title} {\bibinfo {title} {Exchange-enhanced ultrastrong magnon-magnon coupling in a compensated ferrimagnet},\ }\href {https://doi.org/10.1103/PhysRevLett.123.117204} {\bibfield  {journal} {\bibinfo  {journal} {Phys. Rev. Lett.}\ }\textbf {\bibinfo {volume} {123}},\ \bibinfo {pages} {117204} (\bibinfo {year} {2019})}\BibitemShut {NoStop}%
\bibitem [{\citenamefont {Shiota}\ \emph {et~al.}(2020)\citenamefont {Shiota}, \citenamefont {Taniguchi}, \citenamefont {Ishibashi}, \citenamefont {Moriyama},\ and\ \citenamefont {Ono}}]{shiota2020tunable}%
  \BibitemOpen
  \bibfield  {author} {\bibinfo {author} {\bibfnamefont {Y.}~\bibnamefont {Shiota}}, \bibinfo {author} {\bibfnamefont {T.}~\bibnamefont {Taniguchi}}, \bibinfo {author} {\bibfnamefont {M.}~\bibnamefont {Ishibashi}}, \bibinfo {author} {\bibfnamefont {T.}~\bibnamefont {Moriyama}},\ and\ \bibinfo {author} {\bibfnamefont {T.}~\bibnamefont {Ono}},\ }\bibfield  {title} {\bibinfo {title} {Tunable magnon-magnon coupling mediated by dynamic dipolar interaction in synthetic antiferromagnets},\ }\href {https://doi.org/10.1103/PhysRevLett.125.017203} {\bibfield  {journal} {\bibinfo  {journal} {Phys. Rev. Lett.}\ }\textbf {\bibinfo {volume} {125}},\ \bibinfo {pages} {017203} (\bibinfo {year} {2020})}\BibitemShut {NoStop}%
\bibitem [{\citenamefont {Sud}\ \emph {et~al.}(2020)\citenamefont {Sud}, \citenamefont {Zollitsch}, \citenamefont {Kamimaki}, \citenamefont {Dion}, \citenamefont {Khan}, \citenamefont {Iihama}, \citenamefont {Mizukami},\ and\ \citenamefont {Kurebayashi}}]{sud2020tunable}%
  \BibitemOpen
  \bibfield  {author} {\bibinfo {author} {\bibfnamefont {A.}~\bibnamefont {Sud}}, \bibinfo {author} {\bibfnamefont {C.~W.}\ \bibnamefont {Zollitsch}}, \bibinfo {author} {\bibfnamefont {A.}~\bibnamefont {Kamimaki}}, \bibinfo {author} {\bibfnamefont {T.}~\bibnamefont {Dion}}, \bibinfo {author} {\bibfnamefont {S.}~\bibnamefont {Khan}}, \bibinfo {author} {\bibfnamefont {S.}~\bibnamefont {Iihama}}, \bibinfo {author} {\bibfnamefont {S.}~\bibnamefont {Mizukami}},\ and\ \bibinfo {author} {\bibfnamefont {H.}~\bibnamefont {Kurebayashi}},\ }\bibfield  {title} {\bibinfo {title} {Tunable magnon-magnon coupling in synthetic antiferromagnets},\ }\href {https://doi.org/10.1103/PhysRevB.102.100403} {\bibfield  {journal} {\bibinfo  {journal} {Phys. Rev. B}\ }\textbf {\bibinfo {volume} {102}},\ \bibinfo {pages} {100403} (\bibinfo {year} {2020})}\BibitemShut {NoStop}%
\bibitem [{\citenamefont {Wang}\ \emph {et~al.}(2024{\natexlab{a}})\citenamefont {Wang}, \citenamefont {Zhang}, \citenamefont {Li}, \citenamefont {Wei}, \citenamefont {He}, \citenamefont {Xu}, \citenamefont {Xia}, \citenamefont {Luo}, \citenamefont {Li}, \citenamefont {Dong} \emph {et~al.}}]{wang2024ultrastrong}%
  \BibitemOpen
  \bibfield  {author} {\bibinfo {author} {\bibfnamefont {Y.}~\bibnamefont {Wang}}, \bibinfo {author} {\bibfnamefont {Y.}~\bibnamefont {Zhang}}, \bibinfo {author} {\bibfnamefont {C.}~\bibnamefont {Li}}, \bibinfo {author} {\bibfnamefont {J.}~\bibnamefont {Wei}}, \bibinfo {author} {\bibfnamefont {B.}~\bibnamefont {He}}, \bibinfo {author} {\bibfnamefont {H.}~\bibnamefont {Xu}}, \bibinfo {author} {\bibfnamefont {J.}~\bibnamefont {Xia}}, \bibinfo {author} {\bibfnamefont {X.}~\bibnamefont {Luo}}, \bibinfo {author} {\bibfnamefont {J.}~\bibnamefont {Li}}, \bibinfo {author} {\bibfnamefont {J.}~\bibnamefont {Dong}}, \emph {et~al.},\ }\bibfield  {title} {\bibinfo {title} {Ultrastrong to nearly deep-strong magnon-magnon coupling with a high degree of freedom in synthetic antiferromagnets},\ }\href {https://doi.org/10.1038/s41467-024-46474-7} {\bibfield  {journal} {\bibinfo  {journal} {Nature Communications}\ }\textbf {\bibinfo {volume} {15}},\ \bibinfo {pages} {2077} (\bibinfo {year} {2024}{\natexlab{a}})}\BibitemShut
  {NoStop}%
\bibitem [{\citenamefont {Bamba}\ \emph {et~al.}(2022)\citenamefont {Bamba}, \citenamefont {Li}, \citenamefont {Marquez~Peraca},\ and\ \citenamefont {Kono}}]{bamba2022magnonic}%
  \BibitemOpen
  \bibfield  {author} {\bibinfo {author} {\bibfnamefont {M.}~\bibnamefont {Bamba}}, \bibinfo {author} {\bibfnamefont {X.}~\bibnamefont {Li}}, \bibinfo {author} {\bibfnamefont {N.}~\bibnamefont {Marquez~Peraca}},\ and\ \bibinfo {author} {\bibfnamefont {J.}~\bibnamefont {Kono}},\ }\bibfield  {title} {\bibinfo {title} {Magnonic superradiant phase transition},\ }\href {https://doi.org/10.1038/s42005-021-00785-z} {\bibfield  {journal} {\bibinfo  {journal} {Communications Physics}\ }\textbf {\bibinfo {volume} {5}},\ \bibinfo {pages} {1} (\bibinfo {year} {2022})}\BibitemShut {NoStop}%
\bibitem [{\citenamefont {Kim}\ \emph {et~al.}(2024)\citenamefont {Kim}, \citenamefont {Dasgupta}, \citenamefont {Ma}, \citenamefont {Park}, \citenamefont {Wei}, \citenamefont {Luo}, \citenamefont {Doumani}, \citenamefont {Li}, \citenamefont {Yang}, \citenamefont {Cheng}, \citenamefont {Kim}, \citenamefont {Everitt}, \citenamefont {Kimura}, \citenamefont {Nojiri}, \citenamefont {Wang}, \citenamefont {Cao}, \citenamefont {Bamba}, \citenamefont {Hazzard},\ and\ \citenamefont {Kono}}]{kim2024observation}%
  \BibitemOpen
  \bibfield  {author} {\bibinfo {author} {\bibfnamefont {D.}~\bibnamefont {Kim}}, \bibinfo {author} {\bibfnamefont {S.}~\bibnamefont {Dasgupta}}, \bibinfo {author} {\bibfnamefont {X.}~\bibnamefont {Ma}}, \bibinfo {author} {\bibfnamefont {J.-M.}\ \bibnamefont {Park}}, \bibinfo {author} {\bibfnamefont {H.-T.}\ \bibnamefont {Wei}}, \bibinfo {author} {\bibfnamefont {L.}~\bibnamefont {Luo}}, \bibinfo {author} {\bibfnamefont {J.}~\bibnamefont {Doumani}}, \bibinfo {author} {\bibfnamefont {X.}~\bibnamefont {Li}}, \bibinfo {author} {\bibfnamefont {W.}~\bibnamefont {Yang}}, \bibinfo {author} {\bibfnamefont {D.}~\bibnamefont {Cheng}}, \bibinfo {author} {\bibfnamefont {R.~H.~J.}\ \bibnamefont {Kim}}, \bibinfo {author} {\bibfnamefont {H.~O.}\ \bibnamefont {Everitt}}, \bibinfo {author} {\bibfnamefont {S.}~\bibnamefont {Kimura}}, \bibinfo {author} {\bibfnamefont {H.}~\bibnamefont {Nojiri}}, \bibinfo {author} {\bibfnamefont {J.}~\bibnamefont {Wang}}, \bibinfo {author} {\bibfnamefont {S.}~\bibnamefont {Cao}}, \bibinfo
  {author} {\bibfnamefont {M.}~\bibnamefont {Bamba}}, \bibinfo {author} {\bibfnamefont {K.~R.~A.}\ \bibnamefont {Hazzard}},\ and\ \bibinfo {author} {\bibfnamefont {J.}~\bibnamefont {Kono}},\ }\href {https://arxiv.org/abs/2401.01873} {\bibinfo {title} {Observation of the magnonic dicke superradiant phase transition}} (\bibinfo {year} {2024}),\ \Eprint {https://arxiv.org/abs/2401.01873} {arXiv:2401.01873 [quant-ph]} \BibitemShut {NoStop}%
\bibitem [{\citenamefont {Huebl}\ \emph {et~al.}(2013)\citenamefont {Huebl}, \citenamefont {Zollitsch}, \citenamefont {Lotze}, \citenamefont {Hocke}, \citenamefont {Greifenstein}, \citenamefont {Marx}, \citenamefont {Gross},\ and\ \citenamefont {Goennenwein}}]{huebl2013high}%
  \BibitemOpen
  \bibfield  {author} {\bibinfo {author} {\bibfnamefont {H.}~\bibnamefont {Huebl}}, \bibinfo {author} {\bibfnamefont {C.~W.}\ \bibnamefont {Zollitsch}}, \bibinfo {author} {\bibfnamefont {J.}~\bibnamefont {Lotze}}, \bibinfo {author} {\bibfnamefont {F.}~\bibnamefont {Hocke}}, \bibinfo {author} {\bibfnamefont {M.}~\bibnamefont {Greifenstein}}, \bibinfo {author} {\bibfnamefont {A.}~\bibnamefont {Marx}}, \bibinfo {author} {\bibfnamefont {R.}~\bibnamefont {Gross}},\ and\ \bibinfo {author} {\bibfnamefont {S.~T.}\ \bibnamefont {Goennenwein}},\ }\bibfield  {title} {\bibinfo {title} {High cooperativity in coupled microwave resonator ferrimagnetic insulator hybrids},\ }\href {https://doi.org/10.1103/PhysRevLett.111.127003} {\bibfield  {journal} {\bibinfo  {journal} {Physical Review Letters}\ }\textbf {\bibinfo {volume} {111}},\ \bibinfo {pages} {127003} (\bibinfo {year} {2013})}\BibitemShut {NoStop}%
\bibitem [{\citenamefont {Tabuchi}\ \emph {et~al.}(2014)\citenamefont {Tabuchi}, \citenamefont {Ishino}, \citenamefont {Ishikawa}, \citenamefont {Yamazaki}, \citenamefont {Usami},\ and\ \citenamefont {Nakamura}}]{tabuchi2014hybridizing}%
  \BibitemOpen
  \bibfield  {author} {\bibinfo {author} {\bibfnamefont {Y.}~\bibnamefont {Tabuchi}}, \bibinfo {author} {\bibfnamefont {S.}~\bibnamefont {Ishino}}, \bibinfo {author} {\bibfnamefont {T.}~\bibnamefont {Ishikawa}}, \bibinfo {author} {\bibfnamefont {R.}~\bibnamefont {Yamazaki}}, \bibinfo {author} {\bibfnamefont {K.}~\bibnamefont {Usami}},\ and\ \bibinfo {author} {\bibfnamefont {Y.}~\bibnamefont {Nakamura}},\ }\bibfield  {title} {\bibinfo {title} {Hybridizing ferromagnetic magnons and microwave photons in the quantum limit},\ }\href@noop {} {\bibfield  {journal} {\bibinfo  {journal} {Physical review letters}\ }\textbf {\bibinfo {volume} {113}},\ \bibinfo {pages} {083603} (\bibinfo {year} {2014})}\BibitemShut {NoStop}%
\bibitem [{\citenamefont {Wang}\ \emph {et~al.}(2019)\citenamefont {Wang}, \citenamefont {Rao}, \citenamefont {Yang}, \citenamefont {Xu}, \citenamefont {Gui}, \citenamefont {Yao}, \citenamefont {You},\ and\ \citenamefont {Hu}}]{wang2019nonreciprocity}%
  \BibitemOpen
  \bibfield  {author} {\bibinfo {author} {\bibfnamefont {Y.-P.}\ \bibnamefont {Wang}}, \bibinfo {author} {\bibfnamefont {J.}~\bibnamefont {Rao}}, \bibinfo {author} {\bibfnamefont {Y.}~\bibnamefont {Yang}}, \bibinfo {author} {\bibfnamefont {P.-C.}\ \bibnamefont {Xu}}, \bibinfo {author} {\bibfnamefont {Y.}~\bibnamefont {Gui}}, \bibinfo {author} {\bibfnamefont {B.}~\bibnamefont {Yao}}, \bibinfo {author} {\bibfnamefont {J.}~\bibnamefont {You}},\ and\ \bibinfo {author} {\bibfnamefont {C.-M.}\ \bibnamefont {Hu}},\ }\bibfield  {title} {\bibinfo {title} {Nonreciprocity and unidirectional invisibility in cavity magnonics},\ }\href {https://doi.org/10.1103/PhysRevLett.123.127202} {\bibfield  {journal} {\bibinfo  {journal} {Physical review letters}\ }\textbf {\bibinfo {volume} {123}},\ \bibinfo {pages} {127202} (\bibinfo {year} {2019})}\BibitemShut {NoStop}%
\bibitem [{\citenamefont {Lachance-Quirion}\ \emph {et~al.}(2019)\citenamefont {Lachance-Quirion}, \citenamefont {Tabuchi}, \citenamefont {Gloppe}, \citenamefont {Usami},\ and\ \citenamefont {Nakamura}}]{lachance2019hybrid}%
  \BibitemOpen
  \bibfield  {author} {\bibinfo {author} {\bibfnamefont {D.}~\bibnamefont {Lachance-Quirion}}, \bibinfo {author} {\bibfnamefont {Y.}~\bibnamefont {Tabuchi}}, \bibinfo {author} {\bibfnamefont {A.}~\bibnamefont {Gloppe}}, \bibinfo {author} {\bibfnamefont {K.}~\bibnamefont {Usami}},\ and\ \bibinfo {author} {\bibfnamefont {Y.}~\bibnamefont {Nakamura}},\ }\bibfield  {title} {\bibinfo {title} {Hybrid quantum systems based on magnonics},\ }\href@noop {} {\bibfield  {journal} {\bibinfo  {journal} {Applied Physics Express}\ }\textbf {\bibinfo {volume} {12}},\ \bibinfo {pages} {070101} (\bibinfo {year} {2019})}\BibitemShut {NoStop}%
\bibitem [{\citenamefont {Mandal}\ \emph {et~al.}(2020)\citenamefont {Mandal}, \citenamefont {Kapoor}, \citenamefont {Ghosh}, \citenamefont {Jesudasan}, \citenamefont {Manni}, \citenamefont {Thamizhavel}, \citenamefont {Raychaudhuri}, \citenamefont {Singh},\ and\ \citenamefont {Deshmukh}}]{mandal2020coplanar}%
  \BibitemOpen
  \bibfield  {author} {\bibinfo {author} {\bibfnamefont {S.}~\bibnamefont {Mandal}}, \bibinfo {author} {\bibfnamefont {L.~N.}\ \bibnamefont {Kapoor}}, \bibinfo {author} {\bibfnamefont {S.}~\bibnamefont {Ghosh}}, \bibinfo {author} {\bibfnamefont {J.}~\bibnamefont {Jesudasan}}, \bibinfo {author} {\bibfnamefont {S.}~\bibnamefont {Manni}}, \bibinfo {author} {\bibfnamefont {A.}~\bibnamefont {Thamizhavel}}, \bibinfo {author} {\bibfnamefont {P.}~\bibnamefont {Raychaudhuri}}, \bibinfo {author} {\bibfnamefont {V.}~\bibnamefont {Singh}},\ and\ \bibinfo {author} {\bibfnamefont {M.~M.}\ \bibnamefont {Deshmukh}},\ }\bibfield  {title} {\bibinfo {title} {Coplanar cavity for strong coupling between photons and magnons in van der waals antiferromagnet},\ }\bibfield  {journal} {\bibinfo  {journal} {Applied Physics Letters}\ }\textbf {\bibinfo {volume} {117}},\ \href {https://doi.org/10.1063/5.0029112} {10.1063/5.0029112} (\bibinfo {year} {2020})\BibitemShut {NoStop}%
\bibitem [{\citenamefont {Lachance-Quirion}\ \emph {et~al.}(2020)\citenamefont {Lachance-Quirion}, \citenamefont {Wolski}, \citenamefont {Tabuchi}, \citenamefont {Kono}, \citenamefont {Usami},\ and\ \citenamefont {Nakamura}}]{lachance2020entanglement}%
  \BibitemOpen
  \bibfield  {author} {\bibinfo {author} {\bibfnamefont {D.}~\bibnamefont {Lachance-Quirion}}, \bibinfo {author} {\bibfnamefont {S.~P.}\ \bibnamefont {Wolski}}, \bibinfo {author} {\bibfnamefont {Y.}~\bibnamefont {Tabuchi}}, \bibinfo {author} {\bibfnamefont {S.}~\bibnamefont {Kono}}, \bibinfo {author} {\bibfnamefont {K.}~\bibnamefont {Usami}},\ and\ \bibinfo {author} {\bibfnamefont {Y.}~\bibnamefont {Nakamura}},\ }\bibfield  {title} {\bibinfo {title} {Entanglement-based single-shot detection of a single magnon with a superconducting qubit},\ }\href {https://doi.org/10.1126/science.aaz9236} {\bibfield  {journal} {\bibinfo  {journal} {Science}\ }\textbf {\bibinfo {volume} {367}},\ \bibinfo {pages} {425} (\bibinfo {year} {2020})}\BibitemShut {NoStop}%
\bibitem [{\citenamefont {Shim}\ \emph {et~al.}(2020)\citenamefont {Shim}, \citenamefont {Kim}, \citenamefont {Kim},\ and\ \citenamefont {Lee}}]{shim2020enhanced}%
  \BibitemOpen
  \bibfield  {author} {\bibinfo {author} {\bibfnamefont {J.}~\bibnamefont {Shim}}, \bibinfo {author} {\bibfnamefont {S.-J.}\ \bibnamefont {Kim}}, \bibinfo {author} {\bibfnamefont {S.~K.}\ \bibnamefont {Kim}},\ and\ \bibinfo {author} {\bibfnamefont {K.-J.}\ \bibnamefont {Lee}},\ }\bibfield  {title} {\bibinfo {title} {Enhanced magnon-photon coupling at the angular momentum compensation point of ferrimagnets},\ }\href {https://doi.org/10.1103/PhysRevLett.125.027205} {\bibfield  {journal} {\bibinfo  {journal} {Physical review letters}\ }\textbf {\bibinfo {volume} {125}},\ \bibinfo {pages} {027205} (\bibinfo {year} {2020})}\BibitemShut {NoStop}%
\bibitem [{\citenamefont {Zhao}\ \emph {et~al.}(2004)\citenamefont {Zhao}, \citenamefont {Bragas}, \citenamefont {Lockwood},\ and\ \citenamefont {Merlin}}]{zhao2004magnon}%
  \BibitemOpen
  \bibfield  {author} {\bibinfo {author} {\bibfnamefont {J.}~\bibnamefont {Zhao}}, \bibinfo {author} {\bibfnamefont {A.~V.}\ \bibnamefont {Bragas}}, \bibinfo {author} {\bibfnamefont {D.~J.}\ \bibnamefont {Lockwood}},\ and\ \bibinfo {author} {\bibfnamefont {R.}~\bibnamefont {Merlin}},\ }\bibfield  {title} {\bibinfo {title} {Magnon squeezing in an antiferromagnet: Reducing the spin noise below the standard quantum limit},\ }\href {https://doi.org/10.1103/PhysRevLett.93.107203} {\bibfield  {journal} {\bibinfo  {journal} {Phys. Rev. Lett.}\ }\textbf {\bibinfo {volume} {93}},\ \bibinfo {pages} {107203} (\bibinfo {year} {2004})}\BibitemShut {NoStop}%
\bibitem [{\citenamefont {Zhao}\ \emph {et~al.}(2006)\citenamefont {Zhao}, \citenamefont {Bragas}, \citenamefont {Merlin},\ and\ \citenamefont {Lockwood}}]{zhao2006magnon}%
  \BibitemOpen
  \bibfield  {author} {\bibinfo {author} {\bibfnamefont {J.}~\bibnamefont {Zhao}}, \bibinfo {author} {\bibfnamefont {A.~V.}\ \bibnamefont {Bragas}}, \bibinfo {author} {\bibfnamefont {R.}~\bibnamefont {Merlin}},\ and\ \bibinfo {author} {\bibfnamefont {D.~J.}\ \bibnamefont {Lockwood}},\ }\bibfield  {title} {\bibinfo {title} {Magnon squeezing in antiferromagnetic ${\mathrm{mnf}}_{2}$ and ${\mathrm{fef}}_{2}$},\ }\href {https://doi.org/10.1103/PhysRevB.73.184434} {\bibfield  {journal} {\bibinfo  {journal} {Phys. Rev. B}\ }\textbf {\bibinfo {volume} {73}},\ \bibinfo {pages} {184434} (\bibinfo {year} {2006})}\BibitemShut {NoStop}%
\bibitem [{\citenamefont {Kamra}\ and\ \citenamefont {Belzig}(2017)}]{kamra2017spin}%
  \BibitemOpen
  \bibfield  {author} {\bibinfo {author} {\bibfnamefont {A.}~\bibnamefont {Kamra}}\ and\ \bibinfo {author} {\bibfnamefont {W.}~\bibnamefont {Belzig}},\ }\bibfield  {title} {\bibinfo {title} {Spin pumping and shot noise in ferrimagnets: Bridging ferro- and antiferromagnets},\ }\href {https://doi.org/10.1103/PhysRevLett.119.197201} {\bibfield  {journal} {\bibinfo  {journal} {Phys. Rev. Lett.}\ }\textbf {\bibinfo {volume} {119}},\ \bibinfo {pages} {197201} (\bibinfo {year} {2017})}\BibitemShut {NoStop}%
\bibitem [{\citenamefont {Kamra}\ \emph {et~al.}(2019)\citenamefont {Kamra}, \citenamefont {Thingstad}, \citenamefont {Rastelli}, \citenamefont {Duine}, \citenamefont {Brataas}, \citenamefont {Belzig},\ and\ \citenamefont {Sudb\o{}}}]{kamra2019antiferromagnetic}%
  \BibitemOpen
  \bibfield  {author} {\bibinfo {author} {\bibfnamefont {A.}~\bibnamefont {Kamra}}, \bibinfo {author} {\bibfnamefont {E.}~\bibnamefont {Thingstad}}, \bibinfo {author} {\bibfnamefont {G.}~\bibnamefont {Rastelli}}, \bibinfo {author} {\bibfnamefont {R.~A.}\ \bibnamefont {Duine}}, \bibinfo {author} {\bibfnamefont {A.}~\bibnamefont {Brataas}}, \bibinfo {author} {\bibfnamefont {W.}~\bibnamefont {Belzig}},\ and\ \bibinfo {author} {\bibfnamefont {A.}~\bibnamefont {Sudb\o{}}},\ }\bibfield  {title} {\bibinfo {title} {Antiferromagnetic magnons as highly squeezed fock states underlying quantum correlations},\ }\href {https://doi.org/10.1103/PhysRevB.100.174407} {\bibfield  {journal} {\bibinfo  {journal} {Phys. Rev. B}\ }\textbf {\bibinfo {volume} {100}},\ \bibinfo {pages} {174407} (\bibinfo {year} {2019})}\BibitemShut {NoStop}%
\bibitem [{\citenamefont {Kamra}\ \emph {et~al.}(2020)\citenamefont {Kamra}, \citenamefont {Belzig},\ and\ \citenamefont {Brataas}}]{kamra2020magnon}%
  \BibitemOpen
  \bibfield  {author} {\bibinfo {author} {\bibfnamefont {A.}~\bibnamefont {Kamra}}, \bibinfo {author} {\bibfnamefont {W.}~\bibnamefont {Belzig}},\ and\ \bibinfo {author} {\bibfnamefont {A.}~\bibnamefont {Brataas}},\ }\bibfield  {title} {\bibinfo {title} {Magnon-squeezing as a niche of quantum magnonics},\ }\href {https://doi.org/10.1063/5.0021099} {\bibfield  {journal} {\bibinfo  {journal} {Applied Physics Letters}\ }\textbf {\bibinfo {volume} {117}},\ \bibinfo {pages} {090501} (\bibinfo {year} {2020})}\BibitemShut {NoStop}%
\bibitem [{\citenamefont {Zou}\ \emph {et~al.}(2020)\citenamefont {Zou}, \citenamefont {Kim},\ and\ \citenamefont {Tserkovnyak}}]{zou2020tuning}%
  \BibitemOpen
  \bibfield  {author} {\bibinfo {author} {\bibfnamefont {J.}~\bibnamefont {Zou}}, \bibinfo {author} {\bibfnamefont {S.~K.}\ \bibnamefont {Kim}},\ and\ \bibinfo {author} {\bibfnamefont {Y.}~\bibnamefont {Tserkovnyak}},\ }\bibfield  {title} {\bibinfo {title} {Tuning entanglement by squeezing magnons in anisotropic magnets},\ }\href {https://doi.org/10.1103/PhysRevB.101.014416} {\bibfield  {journal} {\bibinfo  {journal} {Physical Review B}\ }\textbf {\bibinfo {volume} {101}},\ \bibinfo {pages} {014416} (\bibinfo {year} {2020})}\BibitemShut {NoStop}%
\bibitem [{\citenamefont {Lee}\ \emph {et~al.}(2023{\natexlab{a}})\citenamefont {Lee}, \citenamefont {Lee},\ and\ \citenamefont {Hwang}}]{lee2023cavity}%
  \BibitemOpen
  \bibfield  {author} {\bibinfo {author} {\bibfnamefont {J.~M.}\ \bibnamefont {Lee}}, \bibinfo {author} {\bibfnamefont {H.-W.}\ \bibnamefont {Lee}},\ and\ \bibinfo {author} {\bibfnamefont {M.-J.}\ \bibnamefont {Hwang}},\ }\bibfield  {title} {\bibinfo {title} {Cavity magnonics with easy-axis ferromagnets: Critically enhanced magnon squeezing and light-matter interaction},\ }\href {https://doi.org/10.1103/PhysRevB.108.L241404} {\bibfield  {journal} {\bibinfo  {journal} {Physical Review B}\ }\textbf {\bibinfo {volume} {108}},\ \bibinfo {pages} {L241404} (\bibinfo {year} {2023}{\natexlab{a}})}\BibitemShut {NoStop}%
\bibitem [{\citenamefont {Lee}\ \emph {et~al.}(2023{\natexlab{b}})\citenamefont {Lee}, \citenamefont {Hwang},\ and\ \citenamefont {Lee}}]{lee2023topological}%
  \BibitemOpen
  \bibfield  {author} {\bibinfo {author} {\bibfnamefont {J.~M.}\ \bibnamefont {Lee}}, \bibinfo {author} {\bibfnamefont {M.-J.}\ \bibnamefont {Hwang}},\ and\ \bibinfo {author} {\bibfnamefont {H.-W.}\ \bibnamefont {Lee}},\ }\bibfield  {title} {\bibinfo {title} {Topological magnon-photon interaction for cavity magnonics},\ }\href {https://doi.org/10.1038/s42005-023-01316-8} {\bibfield  {journal} {\bibinfo  {journal} {Communications Physics}\ }\textbf {\bibinfo {volume} {6}},\ \bibinfo {pages} {194} (\bibinfo {year} {2023}{\natexlab{b}})}\BibitemShut {NoStop}%
\bibitem [{\citenamefont {R\"omling}\ \emph {et~al.}(2023)\citenamefont {R\"omling}, \citenamefont {Vivas-Via\~na}, \citenamefont {Mu\~noz},\ and\ \citenamefont {Kamra}}]{romling2023resolving}%
  \BibitemOpen
  \bibfield  {author} {\bibinfo {author} {\bibfnamefont {A.-L.~E.}\ \bibnamefont {R\"omling}}, \bibinfo {author} {\bibfnamefont {A.}~\bibnamefont {Vivas-Via\~na}}, \bibinfo {author} {\bibfnamefont {C.~S.}\ \bibnamefont {Mu\~noz}},\ and\ \bibinfo {author} {\bibfnamefont {A.}~\bibnamefont {Kamra}},\ }\bibfield  {title} {\bibinfo {title} {Resolving nonclassical magnon composition of a magnetic ground state via a qubit},\ }\href {https://doi.org/10.1103/PhysRevLett.131.143602} {\bibfield  {journal} {\bibinfo  {journal} {Phys. Rev. Lett.}\ }\textbf {\bibinfo {volume} {131}},\ \bibinfo {pages} {143602} (\bibinfo {year} {2023})}\BibitemShut {NoStop}%
\bibitem [{\citenamefont {R\"omling}\ and\ \citenamefont {Kamra}(2024)}]{Romling2024}%
  \BibitemOpen
  \bibfield  {author} {\bibinfo {author} {\bibfnamefont {A.-L.~E.}\ \bibnamefont {R\"omling}}\ and\ \bibinfo {author} {\bibfnamefont {A.}~\bibnamefont {Kamra}},\ }\bibfield  {title} {\bibinfo {title} {Quantum sensing of antiferromagnetic magnon two-mode squeezed vacuum},\ }\href {https://doi.org/10.1103/PhysRevB.109.174410} {\bibfield  {journal} {\bibinfo  {journal} {Phys. Rev. B}\ }\textbf {\bibinfo {volume} {109}},\ \bibinfo {pages} {174410} (\bibinfo {year} {2024})}\BibitemShut {NoStop}%
\bibitem [{\citenamefont {Rom\'an-Roche}\ \emph {et~al.}(2021)\citenamefont {Rom\'an-Roche}, \citenamefont {Luis},\ and\ \citenamefont {Zueco}}]{roman2021photon}%
  \BibitemOpen
  \bibfield  {author} {\bibinfo {author} {\bibfnamefont {J.}~\bibnamefont {Rom\'an-Roche}}, \bibinfo {author} {\bibfnamefont {F.}~\bibnamefont {Luis}},\ and\ \bibinfo {author} {\bibfnamefont {D.}~\bibnamefont {Zueco}},\ }\bibfield  {title} {\bibinfo {title} {Photon condensation and enhanced magnetism in cavity qed},\ }\href {https://doi.org/10.1103/PhysRevLett.127.167201} {\bibfield  {journal} {\bibinfo  {journal} {Phys. Rev. Lett.}\ }\textbf {\bibinfo {volume} {127}},\ \bibinfo {pages} {167201} (\bibinfo {year} {2021})}\BibitemShut {NoStop}%
\bibitem [{\citenamefont {Liu}\ \emph {et~al.}(2023)\citenamefont {Liu}, \citenamefont {Xiong},\ and\ \citenamefont {Ying}}]{liu2023switchable}%
  \BibitemOpen
  \bibfield  {author} {\bibinfo {author} {\bibfnamefont {G.}~\bibnamefont {Liu}}, \bibinfo {author} {\bibfnamefont {W.}~\bibnamefont {Xiong}},\ and\ \bibinfo {author} {\bibfnamefont {Z.-J.}\ \bibnamefont {Ying}},\ }\bibfield  {title} {\bibinfo {title} {Switchable superradiant phase transition with kerr magnons},\ }\href {https://doi.org/10.1103/PhysRevA.108.033704} {\bibfield  {journal} {\bibinfo  {journal} {Phys. Rev. A}\ }\textbf {\bibinfo {volume} {108}},\ \bibinfo {pages} {033704} (\bibinfo {year} {2023})}\BibitemShut {NoStop}%
\bibitem [{\citenamefont {Gr\"unberg}\ \emph {et~al.}(1986)\citenamefont {Gr\"unberg}, \citenamefont {Schreiber}, \citenamefont {Pang}, \citenamefont {Brodsky},\ and\ \citenamefont {Sowers}}]{grunberg1986layered}%
  \BibitemOpen
  \bibfield  {author} {\bibinfo {author} {\bibfnamefont {P.}~\bibnamefont {Gr\"unberg}}, \bibinfo {author} {\bibfnamefont {R.}~\bibnamefont {Schreiber}}, \bibinfo {author} {\bibfnamefont {Y.}~\bibnamefont {Pang}}, \bibinfo {author} {\bibfnamefont {M.~B.}\ \bibnamefont {Brodsky}},\ and\ \bibinfo {author} {\bibfnamefont {H.}~\bibnamefont {Sowers}},\ }\bibfield  {title} {\bibinfo {title} {Layered magnetic structures: Evidence for antiferromagnetic coupling of fe layers across cr interlayers},\ }\href {https://doi.org/10.1103/PhysRevLett.57.2442} {\bibfield  {journal} {\bibinfo  {journal} {Phys. Rev. Lett.}\ }\textbf {\bibinfo {volume} {57}},\ \bibinfo {pages} {2442} (\bibinfo {year} {1986})}\BibitemShut {NoStop}%
\bibitem [{\citenamefont {Parkin}\ \emph {et~al.}(1990)\citenamefont {Parkin}, \citenamefont {More},\ and\ \citenamefont {Roche}}]{parkin1990oscillations}%
  \BibitemOpen
  \bibfield  {author} {\bibinfo {author} {\bibfnamefont {S.~S.~P.}\ \bibnamefont {Parkin}}, \bibinfo {author} {\bibfnamefont {N.}~\bibnamefont {More}},\ and\ \bibinfo {author} {\bibfnamefont {K.~P.}\ \bibnamefont {Roche}},\ }\bibfield  {title} {\bibinfo {title} {Oscillations in exchange coupling and magnetoresistance in metallic superlattice structures: Co/ru, co/cr, and fe/cr},\ }\href {https://doi.org/10.1103/PhysRevLett.64.2304} {\bibfield  {journal} {\bibinfo  {journal} {Phys. Rev. Lett.}\ }\textbf {\bibinfo {volume} {64}},\ \bibinfo {pages} {2304} (\bibinfo {year} {1990})}\BibitemShut {NoStop}%
\bibitem [{\citenamefont {Duine}\ \emph {et~al.}(2018)\citenamefont {Duine}, \citenamefont {Lee}, \citenamefont {Parkin},\ and\ \citenamefont {Stiles}}]{duine2018synthetic}%
  \BibitemOpen
  \bibfield  {author} {\bibinfo {author} {\bibfnamefont {R.}~\bibnamefont {Duine}}, \bibinfo {author} {\bibfnamefont {K.-J.}\ \bibnamefont {Lee}}, \bibinfo {author} {\bibfnamefont {S.~S.}\ \bibnamefont {Parkin}},\ and\ \bibinfo {author} {\bibfnamefont {M.~D.}\ \bibnamefont {Stiles}},\ }\bibfield  {title} {\bibinfo {title} {Synthetic antiferromagnetic spintronics},\ }\href {https://doi.org/10.1038/s41567-018-0050-y} {\bibfield  {journal} {\bibinfo  {journal} {Nature physics}\ }\textbf {\bibinfo {volume} {14}},\ \bibinfo {pages} {217} (\bibinfo {year} {2018})}\BibitemShut {NoStop}%
\bibitem [{\citenamefont {Wuhrer}\ \emph {et~al.}(2022)\citenamefont {Wuhrer}, \citenamefont {Rohling},\ and\ \citenamefont {Belzig}}]{wuhrer2022theory}%
  \BibitemOpen
  \bibfield  {author} {\bibinfo {author} {\bibfnamefont {D.}~\bibnamefont {Wuhrer}}, \bibinfo {author} {\bibfnamefont {N.}~\bibnamefont {Rohling}},\ and\ \bibinfo {author} {\bibfnamefont {W.}~\bibnamefont {Belzig}},\ }\bibfield  {title} {\bibinfo {title} {Theory of quantum entanglement and structure of the two-mode squeezed antiferromagnetic magnon vacuum},\ }\href {https://doi.org/10.1103/PhysRevB.105.054406} {\bibfield  {journal} {\bibinfo  {journal} {Phys. Rev. B}\ }\textbf {\bibinfo {volume} {105}},\ \bibinfo {pages} {054406} (\bibinfo {year} {2022})}\BibitemShut {NoStop}%
\bibitem [{\citenamefont {Rezende}\ \emph {et~al.}(2019)\citenamefont {Rezende}, \citenamefont {Azevedo},\ and\ \citenamefont {Rodr{\'\i}guez-Su{\'a}rez}}]{rezende2019introduction}%
  \BibitemOpen
  \bibfield  {author} {\bibinfo {author} {\bibfnamefont {S.~M.}\ \bibnamefont {Rezende}}, \bibinfo {author} {\bibfnamefont {A.}~\bibnamefont {Azevedo}},\ and\ \bibinfo {author} {\bibfnamefont {R.~L.}\ \bibnamefont {Rodr{\'\i}guez-Su{\'a}rez}},\ }\bibfield  {title} {\bibinfo {title} {Introduction to antiferromagnetic magnons},\ }\bibfield  {journal} {\bibinfo  {journal} {Journal of Applied Physics}\ }\textbf {\bibinfo {volume} {126}},\ \href {https://doi.org/10.1063/1.5109132} {10.1063/1.5109132} (\bibinfo {year} {2019})\BibitemShut {NoStop}%
\bibitem [{\citenamefont {Kamra}\ and\ \citenamefont {Belzig}(2016)}]{kamra2016super}%
  \BibitemOpen
  \bibfield  {author} {\bibinfo {author} {\bibfnamefont {A.}~\bibnamefont {Kamra}}\ and\ \bibinfo {author} {\bibfnamefont {W.}~\bibnamefont {Belzig}},\ }\bibfield  {title} {\bibinfo {title} {Super-poissonian shot noise of squeezed-magnon mediated spin transport},\ }\href {https://doi.org/10.1103/PhysRevLett.116.146601} {\bibfield  {journal} {\bibinfo  {journal} {Phys. Rev. Lett.}\ }\textbf {\bibinfo {volume} {116}},\ \bibinfo {pages} {146601} (\bibinfo {year} {2016})}\BibitemShut {NoStop}%
\bibitem [{\citenamefont {Wang}\ \emph {et~al.}(2024{\natexlab{b}})\citenamefont {Wang}, \citenamefont {Wang}, \citenamefont {Chen}, \citenamefont {Legrand}, \citenamefont {Chen}, \citenamefont {Sheng}, \citenamefont {Xia}, \citenamefont {Lan}, \citenamefont {Zhang}, \citenamefont {Yuan}, \citenamefont {Dong}, \citenamefont {Han}, \citenamefont {Ansermet},\ and\ \citenamefont {Yu}}]{wang2024broad}%
  \BibitemOpen
  \bibfield  {author} {\bibinfo {author} {\bibfnamefont {J.}~\bibnamefont {Wang}}, \bibinfo {author} {\bibfnamefont {H.}~\bibnamefont {Wang}}, \bibinfo {author} {\bibfnamefont {J.}~\bibnamefont {Chen}}, \bibinfo {author} {\bibfnamefont {W.}~\bibnamefont {Legrand}}, \bibinfo {author} {\bibfnamefont {P.}~\bibnamefont {Chen}}, \bibinfo {author} {\bibfnamefont {L.}~\bibnamefont {Sheng}}, \bibinfo {author} {\bibfnamefont {J.}~\bibnamefont {Xia}}, \bibinfo {author} {\bibfnamefont {G.}~\bibnamefont {Lan}}, \bibinfo {author} {\bibfnamefont {Y.}~\bibnamefont {Zhang}}, \bibinfo {author} {\bibfnamefont {R.}~\bibnamefont {Yuan}}, \bibinfo {author} {\bibfnamefont {J.}~\bibnamefont {Dong}}, \bibinfo {author} {\bibfnamefont {X.}~\bibnamefont {Han}}, \bibinfo {author} {\bibfnamefont {J.-P.}\ \bibnamefont {Ansermet}},\ and\ \bibinfo {author} {\bibfnamefont {H.}~\bibnamefont {Yu}},\ }\bibfield  {title} {\bibinfo {title} {Broad-wave-vector spin pumping of flat-band magnons},\ }\href
  {https://doi.org/10.1103/PhysRevApplied.21.044024} {\bibfield  {journal} {\bibinfo  {journal} {Phys. Rev. Appl.}\ }\textbf {\bibinfo {volume} {21}},\ \bibinfo {pages} {044024} (\bibinfo {year} {2024}{\natexlab{b}})}\BibitemShut {NoStop}%
\bibitem [{\citenamefont {Pirandola}\ \emph {et~al.}(2009)\citenamefont {Pirandola}, \citenamefont {Serafini},\ and\ \citenamefont {Lloyd}}]{pirandola2009correlation}%
  \BibitemOpen
  \bibfield  {author} {\bibinfo {author} {\bibfnamefont {S.}~\bibnamefont {Pirandola}}, \bibinfo {author} {\bibfnamefont {A.}~\bibnamefont {Serafini}},\ and\ \bibinfo {author} {\bibfnamefont {S.}~\bibnamefont {Lloyd}},\ }\bibfield  {title} {\bibinfo {title} {Correlation matrices of two-mode bosonic systems},\ }\href {https://doi.org/10.1103/PhysRevA.79.052327} {\bibfield  {journal} {\bibinfo  {journal} {Phys. Rev. A}\ }\textbf {\bibinfo {volume} {79}},\ \bibinfo {pages} {052327} (\bibinfo {year} {2009})}\BibitemShut {NoStop}%
\bibitem [{\citenamefont {Yuan}\ \emph {et~al.}(2020)\citenamefont {Yuan}, \citenamefont {Zheng}, \citenamefont {Ficek}, \citenamefont {He},\ and\ \citenamefont {Yung}}]{yuan2020enhancement}%
  \BibitemOpen
  \bibfield  {author} {\bibinfo {author} {\bibfnamefont {H.~Y.}\ \bibnamefont {Yuan}}, \bibinfo {author} {\bibfnamefont {S.}~\bibnamefont {Zheng}}, \bibinfo {author} {\bibfnamefont {Z.}~\bibnamefont {Ficek}}, \bibinfo {author} {\bibfnamefont {Q.~Y.}\ \bibnamefont {He}},\ and\ \bibinfo {author} {\bibfnamefont {M.-H.}\ \bibnamefont {Yung}},\ }\bibfield  {title} {\bibinfo {title} {Enhancement of magnon-magnon entanglement inside a cavity},\ }\href {https://doi.org/10.1103/PhysRevB.101.014419} {\bibfield  {journal} {\bibinfo  {journal} {Phys. Rev. B}\ }\textbf {\bibinfo {volume} {101}},\ \bibinfo {pages} {014419} (\bibinfo {year} {2020})}\BibitemShut {NoStop}%
\bibitem [{\citenamefont {Shim}\ and\ \citenamefont {Lee}(2022)}]{shim2022enhanced}%
  \BibitemOpen
  \bibfield  {author} {\bibinfo {author} {\bibfnamefont {J.}~\bibnamefont {Shim}}\ and\ \bibinfo {author} {\bibfnamefont {K.-J.}\ \bibnamefont {Lee}},\ }\bibfield  {title} {\bibinfo {title} {Enhanced magnon-magnon entanglement in the vicinity of an angular momentum compensation point of a ferrimagnet in a cavity},\ }\href {https://doi.org/10.1103/PhysRevB.106.L140408} {\bibfield  {journal} {\bibinfo  {journal} {Phys. Rev. B}\ }\textbf {\bibinfo {volume} {106}},\ \bibinfo {pages} {L140408} (\bibinfo {year} {2022})}\BibitemShut {NoStop}%
\bibitem [{\citenamefont {Soykal}\ and\ \citenamefont {Flatt\'e}(2010)}]{soykal2010strong}%
  \BibitemOpen
  \bibfield  {author} {\bibinfo {author} {\bibfnamefont {O.~O.}\ \bibnamefont {Soykal}}\ and\ \bibinfo {author} {\bibfnamefont {M.~E.}\ \bibnamefont {Flatt\'e}},\ }\bibfield  {title} {\bibinfo {title} {Strong field interactions between a nanomagnet and a photonic cavity},\ }\href {https://doi.org/10.1103/PhysRevLett.104.077202} {\bibfield  {journal} {\bibinfo  {journal} {Phys. Rev. Lett.}\ }\textbf {\bibinfo {volume} {104}},\ \bibinfo {pages} {077202} (\bibinfo {year} {2010})}\BibitemShut {NoStop}%
\bibitem [{\citenamefont {Yuan}\ \emph {et~al.}(2022)\citenamefont {Yuan}, \citenamefont {Cao}, \citenamefont {Kamra}, \citenamefont {Duine},\ and\ \citenamefont {Yan}}]{yuan2022quantum}%
  \BibitemOpen
  \bibfield  {author} {\bibinfo {author} {\bibfnamefont {H.}~\bibnamefont {Yuan}}, \bibinfo {author} {\bibfnamefont {Y.}~\bibnamefont {Cao}}, \bibinfo {author} {\bibfnamefont {A.}~\bibnamefont {Kamra}}, \bibinfo {author} {\bibfnamefont {R.~A.}\ \bibnamefont {Duine}},\ and\ \bibinfo {author} {\bibfnamefont {P.}~\bibnamefont {Yan}},\ }\bibfield  {title} {\bibinfo {title} {Quantum magnonics: when magnon spintronics meets quantum information science},\ }\href {https://doi.org/10.1016/j.physrep.2022.03.002} {\bibfield  {journal} {\bibinfo  {journal} {Physics Reports}\ }\textbf {\bibinfo {volume} {965}},\ \bibinfo {pages} {1} (\bibinfo {year} {2022})}\BibitemShut {NoStop}%
\bibitem [{\citenamefont {Harder}\ \emph {et~al.}(2018)\citenamefont {Harder}, \citenamefont {Yang}, \citenamefont {Yao}, \citenamefont {Yu}, \citenamefont {Rao}, \citenamefont {Gui}, \citenamefont {Stamps},\ and\ \citenamefont {Hu}}]{harder2018level}%
  \BibitemOpen
  \bibfield  {author} {\bibinfo {author} {\bibfnamefont {M.}~\bibnamefont {Harder}}, \bibinfo {author} {\bibfnamefont {Y.}~\bibnamefont {Yang}}, \bibinfo {author} {\bibfnamefont {B.}~\bibnamefont {Yao}}, \bibinfo {author} {\bibfnamefont {C.}~\bibnamefont {Yu}}, \bibinfo {author} {\bibfnamefont {J.}~\bibnamefont {Rao}}, \bibinfo {author} {\bibfnamefont {Y.}~\bibnamefont {Gui}}, \bibinfo {author} {\bibfnamefont {R.}~\bibnamefont {Stamps}},\ and\ \bibinfo {author} {\bibfnamefont {C.-M.}\ \bibnamefont {Hu}},\ }\bibfield  {title} {\bibinfo {title} {Level attraction due to dissipative magnon-photon coupling},\ }\href {https://doi.org/10.1103/PhysRevLett.121.137203} {\bibfield  {journal} {\bibinfo  {journal} {Physical review letters}\ }\textbf {\bibinfo {volume} {121}},\ \bibinfo {pages} {137203} (\bibinfo {year} {2018})}\BibitemShut {NoStop}%
\bibitem [{\citenamefont {Zhang}\ and\ \citenamefont {You}(2019)}]{zhang2019higher}%
  \BibitemOpen
  \bibfield  {author} {\bibinfo {author} {\bibfnamefont {G.-Q.}\ \bibnamefont {Zhang}}\ and\ \bibinfo {author} {\bibfnamefont {J.}~\bibnamefont {You}},\ }\bibfield  {title} {\bibinfo {title} {Higher-order exceptional point in a cavity magnonics system},\ }\href {https://doi.org/10.1103/PhysRevB.99.054404} {\bibfield  {journal} {\bibinfo  {journal} {Physical Review B}\ }\textbf {\bibinfo {volume} {99}},\ \bibinfo {pages} {054404} (\bibinfo {year} {2019})}\BibitemShut {NoStop}%
\bibitem [{\citenamefont {Hisatomi}\ \emph {et~al.}(2016)\citenamefont {Hisatomi}, \citenamefont {Osada}, \citenamefont {Tabuchi}, \citenamefont {Ishikawa}, \citenamefont {Noguchi}, \citenamefont {Yamazaki}, \citenamefont {Usami},\ and\ \citenamefont {Nakamura}}]{hisatomi2016bidirectional}%
  \BibitemOpen
  \bibfield  {author} {\bibinfo {author} {\bibfnamefont {R.}~\bibnamefont {Hisatomi}}, \bibinfo {author} {\bibfnamefont {A.}~\bibnamefont {Osada}}, \bibinfo {author} {\bibfnamefont {Y.}~\bibnamefont {Tabuchi}}, \bibinfo {author} {\bibfnamefont {T.}~\bibnamefont {Ishikawa}}, \bibinfo {author} {\bibfnamefont {A.}~\bibnamefont {Noguchi}}, \bibinfo {author} {\bibfnamefont {R.}~\bibnamefont {Yamazaki}}, \bibinfo {author} {\bibfnamefont {K.}~\bibnamefont {Usami}},\ and\ \bibinfo {author} {\bibfnamefont {Y.}~\bibnamefont {Nakamura}},\ }\bibfield  {title} {\bibinfo {title} {Bidirectional conversion between microwave and light via ferromagnetic magnons},\ }\href {https://doi.org/10.1103/PhysRevB.93.174427} {\bibfield  {journal} {\bibinfo  {journal} {Physical Review B}\ }\textbf {\bibinfo {volume} {93}},\ \bibinfo {pages} {174427} (\bibinfo {year} {2016})}\BibitemShut {NoStop}%
\bibitem [{\citenamefont {Bittencourt}\ \emph {et~al.}(2019)\citenamefont {Bittencourt}, \citenamefont {Feulner},\ and\ \citenamefont {Kusminskiy}}]{bittencourt2019magnon}%
  \BibitemOpen
  \bibfield  {author} {\bibinfo {author} {\bibfnamefont {V.~A. S.~V.}\ \bibnamefont {Bittencourt}}, \bibinfo {author} {\bibfnamefont {V.}~\bibnamefont {Feulner}},\ and\ \bibinfo {author} {\bibfnamefont {S.~V.}\ \bibnamefont {Kusminskiy}},\ }\bibfield  {title} {\bibinfo {title} {Magnon heralding in cavity optomagnonics},\ }\href {https://doi.org/10.1103/PhysRevA.100.013810} {\bibfield  {journal} {\bibinfo  {journal} {Phys. Rev. A}\ }\textbf {\bibinfo {volume} {100}},\ \bibinfo {pages} {013810} (\bibinfo {year} {2019})}\BibitemShut {NoStop}%
\bibitem [{\citenamefont {Lambert}\ \emph {et~al.}(2004)\citenamefont {Lambert}, \citenamefont {Emary},\ and\ \citenamefont {Brandes}}]{Lambert2004}%
  \BibitemOpen
  \bibfield  {author} {\bibinfo {author} {\bibfnamefont {N.}~\bibnamefont {Lambert}}, \bibinfo {author} {\bibfnamefont {C.}~\bibnamefont {Emary}},\ and\ \bibinfo {author} {\bibfnamefont {T.}~\bibnamefont {Brandes}},\ }\bibfield  {title} {\bibinfo {title} {Entanglement and the phase transition in single-mode superradiance},\ }\href {https://doi.org/10.1103/PhysRevLett.92.073602} {\bibfield  {journal} {\bibinfo  {journal} {Phys. Rev. Lett.}\ }\textbf {\bibinfo {volume} {92}},\ \bibinfo {pages} {073602} (\bibinfo {year} {2004})}\BibitemShut {NoStop}%
\bibitem [{\citenamefont {Hwang}\ \emph {et~al.}(2015)\citenamefont {Hwang}, \citenamefont {Puebla},\ and\ \citenamefont {Plenio}}]{hwang2015}%
  \BibitemOpen
  \bibfield  {author} {\bibinfo {author} {\bibfnamefont {M.-J.}\ \bibnamefont {Hwang}}, \bibinfo {author} {\bibfnamefont {R.}~\bibnamefont {Puebla}},\ and\ \bibinfo {author} {\bibfnamefont {M.~B.}\ \bibnamefont {Plenio}},\ }\bibfield  {title} {\bibinfo {title} {Quantum phase transition and universal dynamics in the rabi model},\ }\href {https://doi.org/10.1103/PhysRevLett.115.180404} {\bibfield  {journal} {\bibinfo  {journal} {Phys. Rev. Lett.}\ }\textbf {\bibinfo {volume} {115}},\ \bibinfo {pages} {180404} (\bibinfo {year} {2015})}\BibitemShut {NoStop}%
\bibitem [{\citenamefont {Johansson}\ \emph {et~al.}(2012)\citenamefont {Johansson}, \citenamefont {Nation},\ and\ \citenamefont {Nori}}]{johansson2012qutip}%
  \BibitemOpen
  \bibfield  {author} {\bibinfo {author} {\bibfnamefont {J.~R.}\ \bibnamefont {Johansson}}, \bibinfo {author} {\bibfnamefont {P.~D.}\ \bibnamefont {Nation}},\ and\ \bibinfo {author} {\bibfnamefont {F.}~\bibnamefont {Nori}},\ }\bibfield  {title} {\bibinfo {title} {Qutip: An open-source python framework for the dynamics of open quantum systems},\ }\href {https://doi.org/10.1016/j.cpc.2012.02.021} {\bibfield  {journal} {\bibinfo  {journal} {Computer physics communications}\ }\textbf {\bibinfo {volume} {183}},\ \bibinfo {pages} {1760} (\bibinfo {year} {2012})}\BibitemShut {NoStop}%
\end{thebibliography}%

\end{document}